\documentclass[authoryear,3p,nopreprintline]{elsarticle}

\usepackage{amssymb}
\usepackage{graphicx}
\usepackage{microtype}
\usepackage{adjustbox}
\usepackage{amsmath}
\usepackage{booktabs} 
\usepackage{tabularx} 
\usepackage{array}
\usepackage{multirow} 
\usepackage{caption} 
\usepackage{float} 
\usepackage{algorithm}
\usepackage{algorithmic}
\usepackage{natbib} 
\usepackage{enumitem}
\usepackage{subcaption}
\usepackage{listings}
\usepackage{url}
\usepackage{xcolor}
\usepackage{etoolbox}

\usepackage[most]{tcolorbox}
\usepackage{tikz}
\usetikzlibrary{arrows.meta,positioning,fit,backgrounds}

\newbool{highlightorange}
\setbool{highlightorange}{false}

\newcommand{\hlorange}[1]{\ifbool{highlightorange}{\textcolor{orange}{#1}}{#1}}
\newcommand{\CAREtablestyle}{\centering\fontsize{9}{11}\selectfont\setlength{\tabcolsep}{3pt}\renewcommand{\arraystretch}{1.12}}
\usepackage[hidelinks]{hyperref}
\hypersetup{
  pdftitle={CARE: Conversational Affective Response Retrieval for Internet Memes},
  pdfauthor={Jiajun Li, Leihan Zhang, Teng Zhang, Le Zhang, Qiang Yan}
}
\date{November 17, 2026}

\begin{document}

\begin{frontmatter}

\title{CARE: Conversational Affective Response Retrieval for Internet Memes}

\author[bupt]{Jiajun Li}
\author[bupt,keylab]{Leihan Zhang\corref{corresponding}}
\ead{zhangleihan@gmail.com}
\author[bupt]{Teng Zhang}
\author[bistu]{Le Zhang}
\author[bupt,keylab]{Qiang Yan}
\cortext[corresponding]{Corresponding author.}
\address[bupt]{School of Economics and Management, Beijing University of Posts and Telecommunications, No. 10 Xitucheng Road, Haidian District, Beijing 100876, China}
\address[keylab]{Beijing Key  Laboratory  of Multimodal Data  Intelligent Perception  and Governance}
\address[bistu]{College of Computer Science, Beijing Information Science and Technology University, No. 12 Xiaoying East Road, Haidian District, Beijing 100085, China}

\begin{abstract}
Conversational meme retrieval supports response selection in online communication, where relevance depends on both contextual meaning and affective expression. We study this task using two language-specific collections from Douban and Imgflip, containing 98,506 meme records across 5,583 discussion topics. Thread membership supplies weak relevance labels, while likes provide implicit ranking feedback. Automatically extracted sentiment, facial-expression indicators, and cross-modal consistency features support an exploratory analysis of engagement. We propose CARE, a Conversational Affective Response Retrieval framework that combines an emotion-enhanced CLIP retriever with a Jina-based reranker. The retriever incorporates free-form emotion descriptors through gated fusion and learns from multiple positive examples per query. The reranker combines query--image interactions with emotion representations and is trained using engagement-based ranking supervision. Under the reported topic-held-out protocol, CARE achieves Hit@10 of 24.03\% on Douban and 49.33\% on Imgflip, compared with 23.16\% and 41.07\%, respectively, for fine-tuned CLIP. These results support the utility of the complete two-stage system for recovering thread-associated memes. They do not independently establish perceived social resonance or isolate the contribution of affect from reranker training. The study provides a basis for further evaluation of affect-aware retrieval with human relevance judgments and controlled comparisons.
\end{abstract}


\end{frontmatter}

\section{Introduction}\label{intro}
Internet memes are used as responses that express amusement, frustration, agreement, and irony in online communication \citep{shifman2013memes,dynel2016has,yus2021pragmatics}. An intelligent meme retrieval system must identify candidates that fit a textual situation while accounting for how visual and textual cues communicate a stance. Literal image--text similarity alone may miss this response function; conversely, matching emotion without matching context can return inappropriate results.

\begin{figure}[htbp]
\centering
\includegraphics[width=0.85\linewidth]{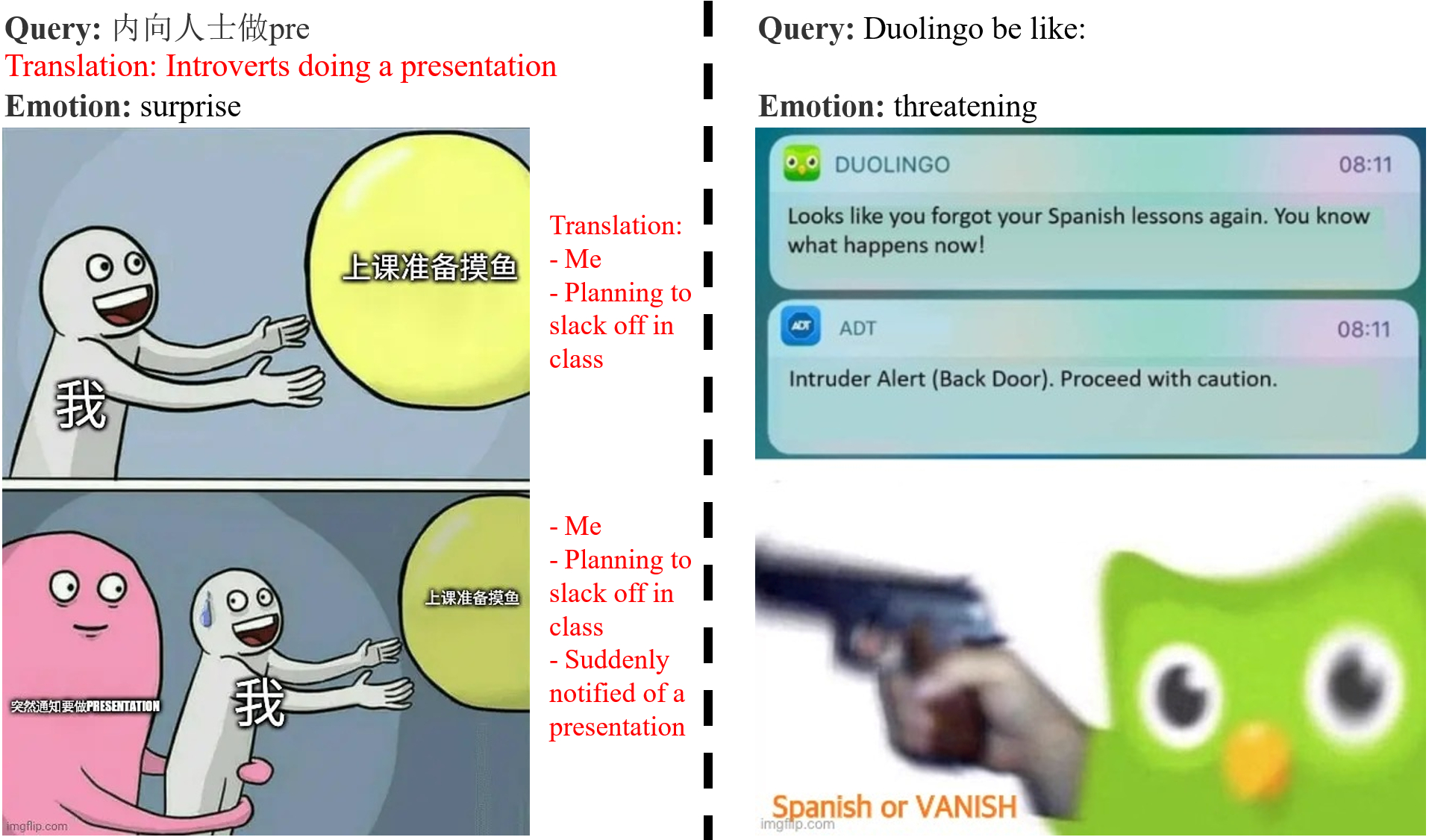}
\caption{Illustrative textual situations and associated memes.}
\label{fig:meme_example}
\end{figure}

Figure~\ref{fig:meme_example} illustrates this challenge. ``Introverts doing a presentation'' invokes anxiety and a change in emotional state, while ``Duolingo be like'' relies on a shared interpretation of persistent reminders. Both cases require contextual interpretation in addition to recognition of depicted objects. These examples motivate affect-aware retrieval, but do not imply that emotion similarity is sufficient for relevance.

Vision--language models such as CLIP provide a useful semantic retrieval backbone \citep{Radford2021LearningTV}. Applying them to conversational memes raises three practical challenges: each context can admit several responses; embedded text and expressive characters can convey meanings beyond visible objects; and naturally occurring interaction records provide incomplete, noisy relevance judgments. Alongside classification benchmarks \citep{sharma-etal-2020-semeval,fersini-etal-2022-semeval}, recent work explicitly addresses meme retrieval and emotion-infused recommendation \citep{le2026memematch,10.1117/12.3065281}. We focus on query--candidate relationships and ranking under thread-based weak supervision.

We investigate these challenges using Chinese and English platform collections. Throughout this paper, affect covers textual sentiment, image-level polarity, and automatically inferred character emotions. We distinguish these observable or model-inferred attributes from perceived social resonance, which is not directly measured. Topic membership and likes provide weak supervision for retrieving thread-associated memes rather than exhaustive judgments of conversational appropriateness.

We propose CARE (Conversational Affective Response Retrieval), a two-stage system that augments a CLIP retriever with emotion descriptors and applies a Jina-based reranker with engagement-based supervision. The exploratory regression examines facial-expression indicators and within-meme semantic and affective consistency, together with sentiment and controls, whereas CARE uses a richer emotion-description vocabulary for retrieval. The regression's semantic and polarity-consistency variables are not explicit alignment objectives in CARE. The contributions are:
\begin{itemize}
\item A conversational meme retrieval collection with 98,506 records across 5,583 topics from Douban and Imgflip, together with automatically extracted semantic and affective attributes and a topic-based evaluation protocol.
\item A two-stage retrieval framework combining gated emotion-feature fusion, multi-positive contrastive learning, and engagement-supervised reranking.
\item An evaluation of the complete system and its reported component configurations on two language-specific collections, accompanied by an exploratory engagement analysis and an explicit account of weak-label and evaluation limitations.
\end{itemize}

\section{Related Work}\label{relatedwork}

This section reviews prior work related to multimodal affect modeling for affect-aware conversational meme retrieval. We first discuss how affect functions in meme communication, drawing on pragmatic and cognitive accounts of meme interpretation. We then review multimodal affect modeling and meme understanding benchmarks, followed by work on conversational and multimodal retrieval. Finally, we position our work by clarifying how it differs from existing approaches.

\subsection{Multimodal Affect in Meme Communication}

Memes are not merely image--text composites but pragmatic acts embedded in digital discourse \citep{shifman2013memes}. Their meaning often emerges through implicature, exaggeration, irony, and shared contextual knowledge \citep{grice1975logic,sperber1986relevance,yus2021pragmatics}. Shifman's framework of content, form, and stance highlights that meme interpretation depends not only on what is represented, but also on how a communicative position is affectively expressed.

Cognitively, meme understanding involves resolving cross-modal incongruity while preserving sufficient coherence for interpretation \citep{yus2021incongruity,tanaka-etal-2022-learning}. Affect is central to this process. Emotional arousal and contagion shape meme propagation \citep{hatfield1993emotional,ayele2023ingredients,beskow2019evolution}, while meme interpretation also depends on contextual cues to stance and emotional orientation \citep{yus2021pragmatics}. Cultural context further modulates affective signaling, particularly in high-context communities where implicit humor and nuanced cues play an important role \citep{zhang2024crosscultural,LI2020102290}.

Together, these studies suggest that multimodal affect provides observable evidence for how users interpret, respond to, and socially resonate with memes. However, most computational approaches still focus on semantic correspondence, leaving the affective basis of conversational meme response largely underexplored.

\subsection{Multimodal Meme Benchmarks and Modeling}

Research on meme understanding has developed from early sentiment-oriented tasks such as Memotion \citep{sharma-etal-2020-semeval} to increasingly multilingual and culturally nuanced benchmarks. A major line of work focuses on harmful content detection, including hate speech and misogyny \citep{suryawanshi-etal-2020-multimodal,fersini-etal-2022-semeval,bui-etal-2025-multi3hate,hossain-etal-2022-mute}. Other benchmarks examine humor and persuasion techniques \citep{xu-etal-2025-punmemecn,dimitrov-etal-2024-semeval}. Unified frameworks such as MemeLens \citep{shahroor2026memelensmultilingualmultitaskvlms} further attempt to consolidate multiple meme understanding tasks.

These benchmarks have substantially advanced meme classification and interpretation, but most are designed around predefined labels rather than conversational response retrieval. As a result, they rarely provide thread-level query--response structures or socially grounded relevance signals needed to study how memes function as affective responses in online interaction.

On the modeling side, vision--language pre-training models such as CLIP \citep{Radford2021LearningTV} and BLIP \citep{li2022blip} establish joint embedding paradigms for image--text alignment. Multimodal large language models (MLLMs), including GPT-4 \citep{openai2024gpt4technicalreport} and Qwen-VL \citep{bai2023qwenvlversatilevisionlanguagemodel}, further improve multimodal reasoning. However, multimodal models can still struggle with irony, metaphor, and culturally specific humor \citep{xu-etal-2025-punmemecn,HE2024103652}.

Recent studies introduce emotional, knowledge-injected, or multi-view reasoning mechanisms to improve meme understanding \citep{shah-etal-2024-memeclip,hu2026multiview}. KID \citep{li2026kidknowledgeinjecteddualheadlearning} grounds harmful meme detection in background knowledge using label-constrained distillation and structured reasoning chains. Its dual-head learning jointly optimizes semantic generation and classification. This work illustrates the importance of contextual knowledge beyond literal image--text correspondence. Its output is a harmfulness prediction, whereas CARE ranks candidate memes for a textual query; the tasks therefore require different supervision and evaluation.

\subsection{Conversational and Multimodal Retrieval}

Pragmatic accounts explain how communicative intent is inferred beyond literal meaning \citep{grice1975logic,sperber1986relevance}. Although these approaches have advanced pragmatic reasoning in language understanding, their connection to multimodal retrieval remains limited. In conversational meme settings, intent is expressed not only through text but also through visual affective tone, character emotions, and cross-modal consistency. This makes meme retrieval closer to response retrieval, where relevance is socially and contextually grounded rather than defined by one-to-one semantic correspondence.

Multimodal retrieval has progressed from shared embedding approaches such as VSE++ \citep{faghri2018vse++} to large-scale contrastive pre-training. Contemporary systems commonly adopt a two-stage architecture: a bi-encoder for efficient coarse retrieval followed by a cross-encoder reranker for fine-grained interaction modeling \citep{jina2024reranker,TIAN2025104165}. This architecture is well suited to conversational meme retrieval, where candidate responses must first be efficiently retrieved and then ranked according to their contextual and affective appropriateness.

For meme-related tasks, prior work improves CLIP through retrieval-guided contrastive learning \citep{mei-etal-2024-improving} or robust adaptation strategies \citep{mei-etal-2025-robust}. More directly related to recommendation, \citet{10.1117/12.3065281} present \textit{A benchmark for cross-modal emotion-infused internet meme recommendation}. This is a relevant antecedent for combining cross-modal retrieval with affective information. We include the associated IEF comparator in the experimental evaluation and do not claim that introducing emotion into meme recommendation is itself new.

MemeMatch \citep{le2026memematch} provides a large-scale meme dataset and retrieval system with nearly one million records from Reddit and Imgflip. It separates local context, including overlaid text and titles, from global visual or template context, and enriches these representations with sentiment, emotion, topic, and usage-intent annotations. It supports retrieval from natural-language or image queries. MemeMatch therefore establishes a direct precedent for context-aware, affect-enriched meme retrieval, rather than only meme classification.

CARE focuses on a thread-associated protocol in which textual main posts supply queries, multiple images from a thread supply weak positives, and engagement supplies ranking feedback. Its evaluated pipeline combines gated emotion descriptors, multi-positive contrastive learning, and supervised reranking. This task and training configuration distinguish the present study's focus from the representation and retrieval capabilities described above; the existing results do not establish superiority over MemeMatch, which has not been evaluated under the same protocol.

\subsection{Positioning of This Work}
CARE studies the use of automatically inferred emotion descriptors in a conversational retrieval pipeline. Relative to MemeMatch \citep{le2026memematch}, KID \citep{li2026kidknowledgeinjecteddualheadlearning}, and the emotion-infused recommendation benchmark \citep{10.1117/12.3065281}, the scope of the contribution is the thread-associated task formulation and the evaluated two-stage training configuration. Its application differs from meme classification because each query can have several relevant candidates and the output is a ranked list. The architectural components draw on established vision--language retrieval, gated fusion, and learning-to-rank methods. 

\section{Dataset Construction and Task Formulation}
\label{dataset}

Modeling multimodal affect for affect-aware conversational meme retrieval requires datasets that preserve both conversational context and response-level social feedback. Classification datasets address tasks such as hate detection, sentiment labeling, and humor recognition, while retrieval-oriented resources such as MemeMatch and the emotion-infused recommendation benchmark explicitly support search or recommendation \citep{le2026memematch,10.1117/12.3065281}. The present collection is organized around discussion topics to study main-post queries, multiple associated memes, and engagement-based ranking.

To study this setting, we construct a large-scale benchmark comprising two language-specific meme collections. The benchmark is designed to support one-to-many conversational meme retrieval, where a main post serves as the query and the associated meme responses form weakly labeled relevance candidates. Engagement signals, especially likes, are used as implicit feedback for ranking thread-associated memes.

\subsection{Benchmark Overview}

The benchmark consists of two independent language-specific collections:

\begin{itemize}
    \item \textbf{Douban (Chinese):} A community-based discussion platform organized around thematic ``Groups,'' characterized by context-rich conversations and culturally implicit humor.
    \item \textbf{Imgflip (English):} A large-scale meme generation and sharing platform supporting image macros and user-generated templates, characterized by high template diversity and rapid meme iteration.
\end{itemize}

\begin{table}[ht]
\CAREtablestyle
  \centering
  \caption{\hlorange{Composition of the conversational meme retrieval benchmark.}}
  \label{tab:dataset_composition}

  \begin{tabularx}{\textwidth}{lXXX}
  \toprule
  \textbf{Dataset} & \textbf{\#Topics} & \textbf{\#Memes} & \textbf{\#Memes/\#Topics} \\
  \midrule
  Douban & 1,836 & 13,387 & 7.29 \\
  Imgflip & 3,747 & 85,119 & 22.72 \\
  \midrule
  \textbf{Overall} & \textbf{5,583} & \textbf{98,506} & \textbf{17.64} \\
  \bottomrule
  \end{tabularx}

\end{table}

After preprocessing, the final benchmark contains 5,583 topics and 98,506 multimodal meme records. The detailed composition of the two sub-datasets is summarized in Table~\ref{tab:dataset_composition}. Unlike template-centric meme datasets, our benchmark is organized around conversational threads, enabling retrieval scenarios in which a single textual situation may elicit multiple meme responses.

\subsection{Data Collection and Preprocessing}

\paragraph{Data Sources}

\textbf{Douban (Chinese).} Data was collected from June 1, 2023 to September 1, 2023 across six meme-oriented communities. The raw crawl included post topics, textual content, comments, images, and engagement metrics (likes and favorites), resulting in 68,943 topics and 819,507 comments and 331,431 meme images.

\textbf{Imgflip (English).} Data was collected up to April 2025 across diverse meme streams. The raw corpus included 125,124 topics, over 1.79 million comments, and 312,914 candidate meme images.

\paragraph{Preprocessing and Quality Control}

To ensure multimodal validity and sufficient conversational context, we adopt a standardized preprocessing protocol. Corrupted text, special characters, duplicate topics, and duplicate images are removed, and posts without textual content are excluded.

To preserve conversational richness, only threads with a minimum number of comments are retained (Douban: $\geq$10; Imgflip: $\geq$2). Low-engagement meme responses are further filtered on Imgflip using a threshold of $\geq$2 likes. These thresholds are determined based on the scale of the raw corpus and the resulting dataset size after filtering, balancing conversational depth, engagement representativeness, and sample sufficiency for stable empirical analysis and retrieval evaluation.

Images are normalized to consistent formats, and incomplete or low-quality samples are discarded.

\begin{figure}[htbp]
\centering
\begin{adjustbox}{max totalsize={\textwidth}{0.8\textheight},center}
\begin{minipage}{\textwidth}
\begin{minipage}{0.44\linewidth}
\centering
\textbf{(a) Raw conversational records}

\vspace{0.5em}

\begin{tcolorbox}[
    colback=white,
    colframe=black!50,
    boxrule=0.5pt,
    arc=1mm,
    left=3pt,
    right=3pt,
    top=3pt,
    bottom=3pt,
    fontupper=\footnotesize,
    title={Raw record A: 9hhyh4.jpg}
]
\centering
\includegraphics[width=0.78\linewidth]{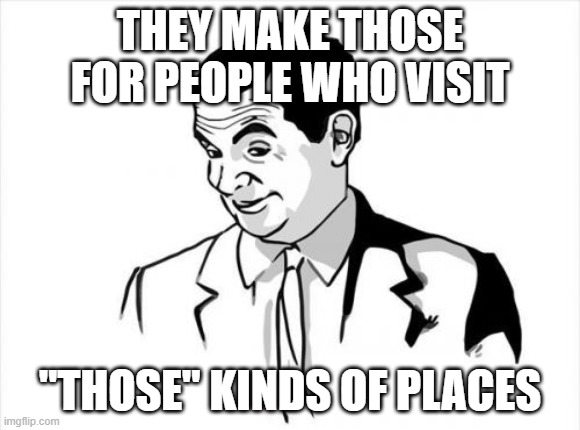}

\vspace{0.4em}
\raggedright
\textbf{Text:} Seriously tho, where else would you put your phone?\\
\textbf{Type:} Source post\\
\textbf{Alt text:} If You Know What I Mean Bean Meme\ldots\\
\textbf{Likes:} 4
\end{tcolorbox}

\vspace{0.7em}

\begin{tcolorbox}[
    colback=white,
    colframe=black!50,
    boxrule=0.5pt,
    arc=1mm,
    left=3pt,
    right=3pt,
    top=3pt,
    bottom=3pt,
    fontupper=\footnotesize,
    title={Raw record B: 85zozy.jpg}
]
\centering
\includegraphics[width=0.78\linewidth]{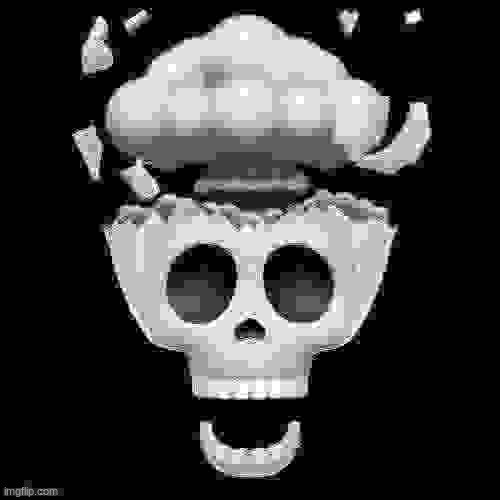}

\vspace{0.4em}
\raggedright
\textbf{Text:} And it's a child too.\\
\textbf{Type:} Comment response\\
\textbf{Alt text:} Man I'm dead\ldots\\
\textbf{Likes:} 2
\end{tcolorbox}

\end{minipage}
\hfill
\begin{minipage}{0.52\linewidth}
\centering
\textbf{(b) Automatic affective representation pipeline}

\vspace{0.5em}

\begin{tikzpicture}[
    node distance=0.50cm,
    box/.style={
        rectangle,
        rounded corners,
        draw,
        align=center,
        minimum width=5.2cm,
        minimum height=0.70cm,
        font=\footnotesize
    },
    arrow/.style={-{Latex[length=2mm]}, thick}
]
\node[box] (raw) {Raw conversational meme};
\node[box, below=of raw] (desc) {Visual description + OCR text};
\node[box, below=of desc] (sent) {Text / image sentiment};
\node[box, below=of sent] (emo) {Emotion descriptors +\\facial-expression labels};
\node[box, below=of emo] (cons) {Semantic consistency + affective consistency};
\node[box, below=of cons] (eng) {Engagement feedback};
\node[box, below=of eng] (rep) {Structured affective representation};

\draw[arrow] (raw) -- (desc);
\draw[arrow] (desc) -- (sent);
\draw[arrow] (sent) -- (emo);
\draw[arrow] (emo) -- (cons);
\draw[arrow] (cons) -- (eng);
\draw[arrow] (eng) -- (rep);
\end{tikzpicture}

\vspace{0.8em}

\textbf{(c) Example extracted affective records}

\vspace{0.4em}

\begin{tcolorbox}[
    colback=white,
    colframe=black!50,
    boxrule=0.5pt,
    arc=1mm,
    left=3pt,
    right=3pt,
    top=3pt,
    bottom=3pt,
    fontupper=\footnotesize,
    title={Structured record A}
]
\textbf{Visual description:} Black-and-white cartoon of a person in a suit and tie with overlaid text.\\
\textbf{OCR:} THEY MAKE THOSE FOR PEOPLE WHO VISIT ``THOSE'' KINDS OF PLACES\\
\textbf{Sentiment:} image = positive; text = positive\\
\textbf{Emotion:} Humor; Happy = 1\\
\textbf{Consistency:} semantic = 0.5268; affective = 1\\
\textbf{Engagement:} $E=\log(4+1)$
\end{tcolorbox}

\vspace{0.55em}

\begin{tcolorbox}[
    colback=white,
    colframe=black!50,
    boxrule=0.5pt,
    arc=1mm,
    left=3pt,
    right=3pt,
    top=3pt,
    bottom=3pt,
    fontupper=\footnotesize,
    title={Structured record B}
]
\textbf{Visual description:} Cartoon skull with the top of the head blown open, showing the brain.\\
\textbf{OCR:} --\\
\textbf{Sentiment:} image = negative; text = negative\\
\textbf{Emotion:} Shock; Surprise = 1\\
\textbf{Consistency:} semantic = 0.5214; affective = 1\\
\textbf{Engagement:} $E=\log(2+1)$
\end{tcolorbox}

\end{minipage}

\vspace{0.5em}

\begin{minipage}{0.95\linewidth}
\footnotesize
\textit{Note: Raw conversational meme records are transformed into structured multimodal affective representations by extracting visual descriptions, OCR text, sentiment polarity, emotion descriptors, facial-expression indicators, cross-modal consistency scores, and engagement feedback.}
\end{minipage}
\end{minipage}
\end{adjustbox}
\caption{Illustrative construction of structured multimodal affective records from raw conversational memes. Free-form emotion descriptions (e.g., Humor and Shock) support retrieval, whereas facial-expression indicators (e.g., Happy = 1 and Surprise = 1) support regression analysis.}
\label{fig:dataset_sample}
\end{figure}

After preprocessing and filtering for valid multimodal availability, the final dataset comprises \textbf{1,836} topics and \textbf{13,387} multimodal meme records for Douban, and \textbf{3,747} topics with \textbf{85,119} multimodal meme records for Imgflip.

Semantic and affective annotations are extracted automatically using platform-specific vision--language models, and consistency measures are derived from these annotations for empirical analysis and retrieval experiments. Representative examples of raw metadata and corresponding automatically extracted multimodal annotations are presented in Figure~\ref{fig:dataset_sample}. These examples illustrate how conversational context, OCR text, visual descriptions, affective attributes, and engagement feedback are jointly structured into retrieval-ready records.

\subsection{Task Formulation}

We formulate the task as \textbf{affect-aware conversational meme retrieval}. Given a textual query $q_i$ corresponding to the main post of a discussion thread, the objective is to retrieve meme responses that are appropriate to the textual situation and its affective expression.

Unlike conventional one-to-one image--text retrieval, this task is inherently open-ended: a single conversational context may elicit multiple acceptable meme responses that differ in humor style, affective stance, and pragmatic expression. Therefore, relevance is not defined as exact semantic correspondence, but as observed membership in the same conversational thread.

Formally, let $Q=\{q_i\}_{i=1}^{|Q|}$ denote the set of textual queries. Each query $q_i$ is associated with a source meme image $m_{q_i}$ in the main post and a set of comment meme responses $M_{q_i}=\{m_{q_i,j}\}_{j=1}^{N_i}$ in the corresponding thread, where $N_i$ denotes the number of valid meme responses. Let $M$ denote the collection of all meme images. The retrieval task is to return the top-$k$ meme responses from $M$ for a given query $q_i$.

The ordered relevance list for $q_i$ is defined as follows. The source meme $m_{q_i}$ is placed first because it provides the original visual context of the conversation. The remaining comment memes in $M_{q_i}$ are ranked in descending order of their number of likes, which serve as implicit engagement feedback within the thread.

\begin{figure}[ht]
  \centering
  \includegraphics[width=0.7\linewidth]{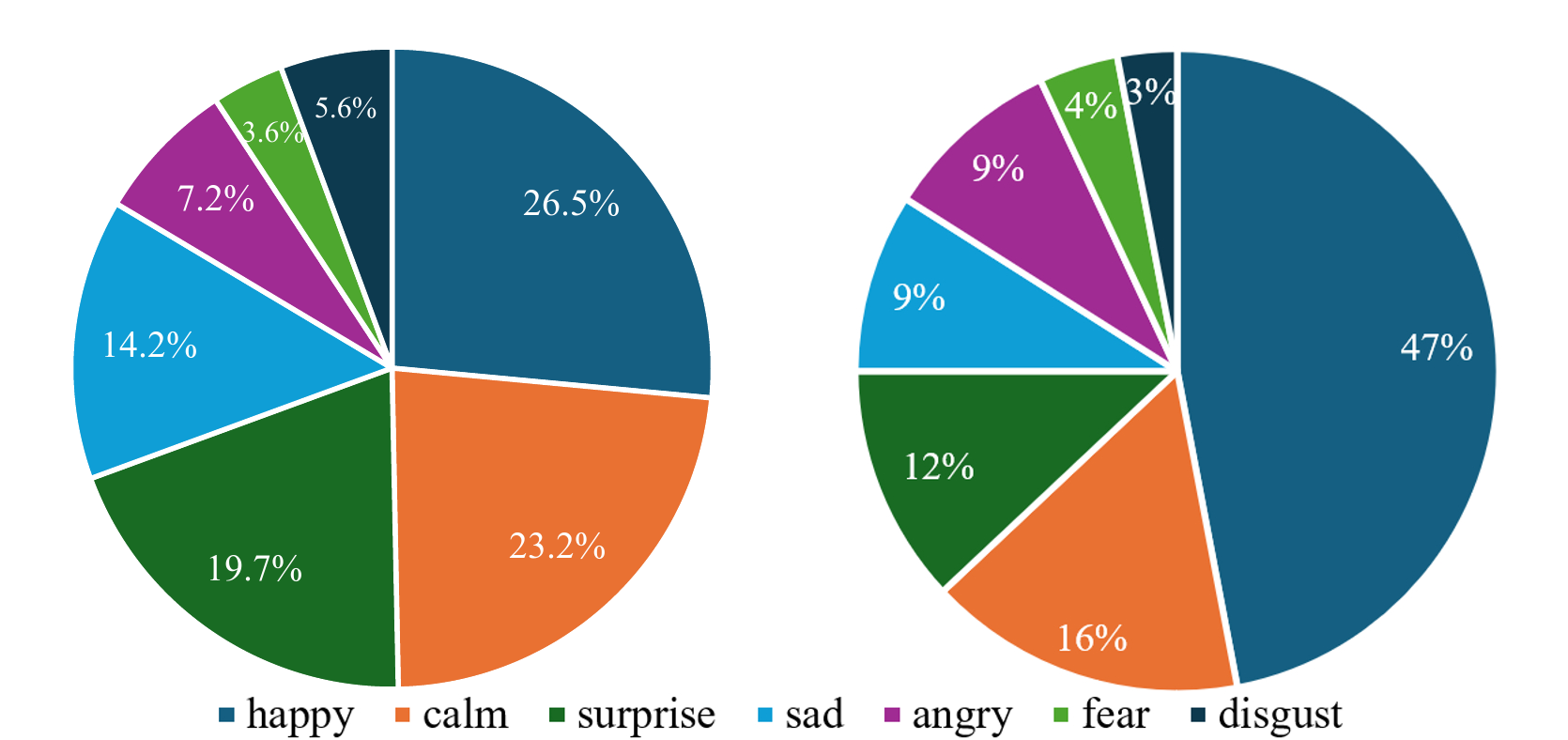}
  \caption{\hlorange{Distribution of facial-expression labels in the Douban (CN; left) and Imgflip (EN; right) datasets. Percentages are based on label occurrences within each dataset.}}
  \label{fig:emotion_dist}
\end{figure}

\subsection{Feature Engineering and Operationalization}
\label{sec:feature_engineering}

To transform raw conversational meme records into structured variables, we implement an automated annotation and representation pipeline that extracts multimodal affective and semantic signals for empirical analysis and response ranking.

\subsubsection{Automated Multimodal Annotation}

We use \textbf{Doubao Vision Pro} for the Douban collection and \textbf{Gemini Pro} \citep{team2023gemini} for the Imgflip collection. Each platform-specific model performs OCR and produces visual descriptions, sentiment labels, free-form emotion descriptors, and facial-expression labels in the same zero-shot annotation pass. The multimodal annotation prompt is provided in~\ref{app:llm_prompts}. The extracted attributes include:

\begin{itemize}

    \item \textbf{Textual Sentiment:} sentiment polarity expressed in meme text (positive / neutral / negative);

    \item \textbf{Image-level Sentiment:} overall affective tone conveyed by the visual content;

    \item \textbf{Emotion Descriptions and Facial Expressions:} free-form emotion descriptors for retrieval and facial-expression labels for regression, with the latter drawn from seven categories (happy, calm, angry, sad, fear, disgust, surprise);

    \item \textbf{Visual Description and OCR Text:} textual representations of visual content and embedded meme captions for semantic consistency estimation.

\end{itemize}

In this study, \textit{affect} is an umbrella term for multimodal emotional signals, and \textit{sentiment} denotes textual or image-level polarity. The same multimodal annotation pass produces two distinct character-level fields. For regression, seven facial-expression labels are converted into meme-level binary indicators: each indicator equals 1 if the corresponding facial-expression category occurs for at least one character in the meme, and 0 otherwise. For retrieval, the original free-form emotion descriptors provide richer affective information, including reactions inferred from the full meme context, and are not restricted to these seven facial-expression categories.

\subsubsection{Semantic Representation and Consistency}

To quantify cross-modal semantic alignment, we encode the textual content of each meme (including OCR text when available) and its automatically generated visual description using the \textbf{BGE-M3} embedding model \citep{bge-m3}. Let $\mathbf{v}_T$ and $\mathbf{v}_V$ denote the text embedding vectors of the meme text and generated visual description, respectively.

Cross-modal semantic consistency is computed as the cosine similarity between the two representations:

\[
S_{\mathrm{sem}}(T,V)=
\frac{\mathbf{v}_T\cdot\mathbf{v}_V}
{\|\mathbf{v}_T\|\|\mathbf{v}_V\|}.
\]

Higher values indicate stronger agreement between the meme text and its generated visual description. This is a description-mediated proxy for cross-modal consistency, not a direct comparison of text and image embeddings.

To capture affective alignment, we define an affective consistency indicator based on the agreement between textual sentiment and image-level sentiment. Specifically,
\[
S_{\mathrm{aff}}=
\begin{cases}
1, & \text{if the two sentiment polarities are identical},\\
0, & \text{otherwise}.
\end{cases}
\]

Unlike semantic consistency, which measures conceptual correspondence across modalities, affective consistency reflects whether the two modalities convey a coherent emotional orientation. These variables summarize agreement within a meme; they do not directly measure compatibility between a query and a response. Polarity disagreement can also support appropriate irony or sarcasm.

\subsubsection{Variable Operationalization}

\begin{table}[htbp]
\CAREtablestyle
\centering
\caption{Construct definitions and operationalization.}
\label{tab:variables}

\begin{tabularx}{\textwidth}{lXl}
\toprule
\textbf{Construct} & \textbf{Operationalization} & \textbf{Type} \\
\midrule
\textit{Content Length} & Character count of the meme textual description & Continuous \\
\textit{Textual Sentiment} & Sentiment polarity expressed in meme text (1 = positive, 0 = neutral, -1 = negative) & Categorical \\
\textit{Visual Sentiment} & Sentiment polarity of the image-level affective tone (1 = positive, 0 = neutral, -1 = negative) & Categorical \\
\textit{Facial Expression} & Meme-level binary indicators for seven facial-expression categories & Binary \\
\textit{Semantic Consistency} & Cosine similarity between embeddings of meme text and a generated visual description & Continuous \\
\textit{Affective Consistency} & Binary indicator of sentiment polarity agreement between text and image & Binary \\
\textit{Topic Popularity} & $\log(\textit{Favorites} + 1)$; available only for Douban & Continuous \\
\textit{Engagement ($E_i$)} & $\log(\textit{Likes}_i + 1)$, used as engagement-based implicit feedback & Continuous \\
\bottomrule
\end{tabularx}
\end{table}

To examine associations between multimodal affect and engagement, we operationalize variables covering textual affect, visual affect, cross-modal consistency, and engagement-based social feedback. Table~\ref{tab:variables} summarizes the construct definitions.

In this study, likes are used as implicit social feedback rather than explicit relevance labels. We therefore define the engagement score of meme response $E_i$ as:

\begin{equation}
E_i = \log\left(\textit{Likes}_i + 1\right),
\label{eq:social_resonance}
\end{equation}

where $E_i$ denotes the log-transformed engagement level of meme response $m_i$. Semantic consistency captures cross-modal conceptual coherence, affective consistency captures polarity-level affective alignment, and facial-expression variables represent the seven automatically inferred facial-expression categories used in regression.

\subsubsection{Descriptive Statistics}

\begin{table}[htbp]
\CAREtablestyle
\centering
\caption{Comparative descriptive statistics for Douban and Imgflip.}
\label{tab:descriptive_stats}
\begin{tabular}{lcccccccccc}
\toprule
 & \multicolumn{5}{c}{\textbf{Douban (Chinese)}} & \multicolumn{5}{c}{\textbf{Imgflip (English)}} \\
\cmidrule(lr){2-6} \cmidrule(lr){7-11}
\textbf{Variable} & \textbf{Mean} & \textbf{SD} & \textbf{Min} & \textbf{Median} & \textbf{Max} & \textbf{Mean} & \textbf{SD} & \textbf{Min} & \textbf{Median} & \textbf{Max} \\
\midrule
\multicolumn{11}{l}{\textbf{\textit{Dependent Variable}}} \\
\textit{Engagement} & 1.56 & 1.55 & 0.00 & 1.10 & 8.64 & 1.43 & 0.37 & 1.10 & 1.39 & 5.88 \\
\midrule
\multicolumn{11}{l}{\textbf{\textit{Independent Variables}}} \\
\textit{Textual Sentiment} & -0.40 & 0.67 & -1.00 & -1.00 & 1.00 & 0.22 & 0.80 & -1.00 & 0.00 & 1.00 \\
\textit{Visual Sentiment} & 0.22 & 0.94 & -1.00 & 1.00 & 1.00 & 0.25 & 0.81 & -1.00 & 0.00 & 1.00 \\
\textit{Happy} & 0.34 & 0.47 & 0.00 & 0.00 & 1.00 & 0.44 & 0.50 & 0.00 & 0.00 & 1.00 \\
\textit{Calm} & 0.29 & 0.46 & 0.00 & 0.00 & 1.00 & 0.15 & 0.36 & 0.00 & 0.00 & 1.00 \\
\textit{Angry} & 0.09 & 0.29 & 0.00 & 0.00 & 1.00 & 0.08 & 0.27 & 0.00 & 0.00 & 1.00 \\
\textit{Sad} & 0.18 & 0.38 & 0.00 & 0.00 & 1.00 & 0.09 & 0.28 & 0.00 & 0.00 & 1.00 \\
\textit{Fear} & 0.05 & 0.21 & 0.00 & 0.00 & 1.00 & 0.03 & 0.18 & 0.00 & 0.00 & 1.00 \\
\textit{Disgust} & 0.07 & 0.26 & 0.00 & 0.00 & 1.00 & 0.03 & 0.16 & 0.00 & 0.00 & 1.00 \\
\textit{Surprise} & 0.25 & 0.43 & 0.00 & 0.00 & 1.00 & 0.11 & 0.32 & 0.00 & 0.00 & 1.00 \\
\textit{Semantic Consistency} & 0.51 & 0.09 & 0.24 & 0.51 & 0.93 & 0.52 & 0.10 & 0.16 & 0.52 & 0.95 \\
\textit{Affective Consistency} & 0.50 & 0.50 & 0.00 & 1.00 & 1.00 & 0.82 & 0.38 & 0.00 & 1.00 & 1.00 \\
\midrule
\multicolumn{11}{l}{\textbf{\textit{Control Variables}}} \\
\textit{Content Length} & 15.72 & 7.50 & 1.00 & 14.00 & 51.00 & 94.90585 & 52.33017 & 5 & 84 & 1853 \\
\textit{Topic Popularity} & 2.44 & 1.72 & 0.00 & 2.30 & 7.68 & --- & --- & --- & --- & --- \\
\bottomrule
\multicolumn{11}{l}{\textit{Note: Engagement and Topic Popularity are log-transformed like and favorite counts, respectively.}} \\
\multicolumn{11}{l}{\textit{Topic Popularity is available only for the Douban dataset.}} \\
\end{tabular}

\end{table}

Table~\ref{tab:descriptive_stats} reports descriptive statistics for the Chinese and English subsets. The Douban Engagement statistics are based on 13,387 records, with SD denoting the sample standard deviation. Semantic consistency shows similar distributions across platforms (mean $\approx 0.52$), whereas engagement levels and affective distributions differ across language environments. In Douban, Happy, Calm, and Surprise account for 26.48\%, 23.21\%, and 19.72\% of facial-expression labels, respectively (Figure~\ref{fig:emotion_dist}), showing the prominence of these expressions in the collection. These observations motivate the subsequent analysis of multimodal affect and engagement.

\section{Exploratory Analysis of Affect and Engagement}\label{theory}
We examine the contributions of facial-expression indicators and within-meme semantic and affective consistency to explaining engagement, alongside sentiment and control variables. This exploratory analysis motivates affect-aware retrieval but does not exhaust the richer emotion-description vocabulary used by CARE. Communication research distinguishes literal content from contextual and pragmatic interpretation \citep{morris1938foundations,levinson1983pragmatics,grice1975logic,sperber1986relevance}. This motivates examining both individual affective attributes and within-meme consistency, without treating either as a direct measurement of response appropriateness.

The analysis asks whether affective variables and cross-modal consistency explain additional variation in log-transformed likes beyond the available controls. Likes are an engagement proxy affected by exposure, posting time, audience, and platform mechanisms. 

\subsection{Model Specification}

We estimate blockwise linear regression models predicting engagement ($E_i$). As defined in Eq.~\ref{eq:social_resonance}, $E_i$ is computed from log-transformed likes and is treated as an engagement outcome. The explanatory variables are organized into three conceptual blocks:

\begin{itemize}

    \item \textbf{Control Variables ($\mathbf{X}_{control}$):} Content Length and Topic Popularity (available for Douban only).

    \item \textbf{Multimodal Affect Signals ($\mathbf{X}_{affect}$):} Textual Sentiment, Visual Sentiment, and Facial Expression indicators.

    \item \textbf{Cross-modal Alignment ($\mathbf{X}_{align}$):} Semantic Consistency and Affective Consistency.

\end{itemize}

The estimation model is specified as:

\begin{equation}
E_i =
\beta_0
+\boldsymbol{\beta}_{control}^{\top}\mathbf{X}_{control,i}
+\boldsymbol{\beta}_{affect}^{\top}\mathbf{X}_{affect,i}
+\boldsymbol{\beta}_{align}^{\top}\mathbf{X}_{align,i}
+\epsilon_i .
\label{eq:regression_model}
\end{equation}

The blocks are entered sequentially: controls, sentiment and facial-expression indicators, semantic consistency, and affective consistency. Here, hierarchical refers to sequential block entry, not a multilevel model. The displayed specification contains no topic fixed effects or random intercepts. Differences in adjusted $R^2$ describe incremental fit rather than causal contributions.

\subsection{Regression Results}

\begin{table}[htbp]
\CAREtablestyle
  \caption{Blockwise regression results for the Douban dataset.}
  \label{tab:douban_full_regression}
\begin{tabular}{lcccc}
  \toprule
  \textbf{Variables} & \textbf{Model 1} & \textbf{Model 2} & \textbf{Model 3} & \textbf{Model 4} \\
  \midrule
  (Constant) & $1.1453^{***}$ & $0.9915^{***}$ & $0.5079^{***}$ & $0.3945^{***}$ \\
  & $(0.02640)$ & $(0.04017)$ & $(0.08342)$ & $(0.08405)$ \\
  \midrule
  \textbf{\textit{Control Variables}} & & & & \\
  \quad \textit{Content Length} & $-0.0021^{***}$ & $-0.0020^{***}$ & $-0.0021^{***}$ & $-0.0021^{***}$ \\
  & $(0.00037)$ & $(0.00037)$ & $(0.00037)$ & $(0.00037)$ \\
  \quad \textit{Topic Popularity} & $0.2016^{***}$ & $0.2009^{***}$ & $0.2062^{***}$ & $0.2047^{***}$ \\
  & $(0.00761)$ & $(0.00763)$ & $(0.00766)$ & $(0.00764)$ \\
  \midrule
  \textbf{\textit{Affective Signals}} & & & & \\
  \quad \textit{Textual Sentiment} &  & $0.0705^{***}$ & $0.0732^{***}$ & $0.0727^{***}$ \\
  &  & $(0.01961)$ & $(0.01958)$ & $(0.01952)$ \\
  \quad \textit{Visual Sentiment} &  & $0.0004$ & $-0.0022$ & $-0.0014$ \\
  &  & $(0.01531)$ & $(0.01529)$ & $(0.01525)$ \\
  \quad \textit{Happy} &  & $0.0695^{*}$ & $0.0594$ & $0.0590$ \\
  &  & $(0.03376)$ & $(0.03374)$ & $(0.03364)$ \\
  \quad \textit{Calm} &  & $0.0959^{**}$ & $0.0902^{**}$ & $0.0930^{**}$ \\
  &  & $(0.03237)$ & $(0.03233)$ & $(0.03223)$ \\
  \quad \textit{Angry} &  & $0.1565^{**}$ & $0.1498^{**}$ & $0.1530^{**}$ \\
  &  & $(0.04873)$ & $(0.04866)$ & $(0.04851)$ \\
  \quad \textit{Sad} &  & $-0.0061$ & $-0.0256$ & $-0.0198$ \\
  &  & $(0.03955)$ & $(0.03960)$ & $(0.03948)$ \\
  \quad \textit{Fear} &  & $0.0586$ & $0.0492$ & $0.0513$ \\
  &  & $(0.06373)$ & $(0.06365)$ & $(0.06344)$ \\
  \quad \textit{Disgust} &  & $0.3344^{***}$ & $0.3138^{***}$ & $0.3128^{***}$ \\
  &  & $(0.05300)$ & $(0.05301)$ & $(0.05284)$ \\
  \quad \textit{Surprise} &  & $0.1421^{***}$ & $0.1316^{***}$ & $0.1308^{***}$ \\
  &  & $(0.03373)$ & $(0.03371)$ & $(0.03360)$ \\
  \midrule
  \textbf{\textit{Cross-modal Consistency}} & & & & \\
  \quad \textit{Semantic Consistency} &  &  & $0.9498^{***}$ & $0.9354^{***}$ \\
  &  &  & $(0.14366)$ & $(0.14322)$ \\
  \quad \textit{Affective Consistency} &  &  &  & $0.2414^{***}$ \\
  &  &  &  & $(0.02599)$ \\
  \midrule
  \textbf{\textit{N}} & 13,387 & 13,387 & 13,387 & 13,387 \\
  \textbf{\textit{$R^2$}} & 0.052023 & 0.057432 & 0.060502 & 0.066525 \\
  \textbf{\textit{Adjusted $R^2$}} & 0.051882 & 0.056656 & 0.059659 & 0.065618 \\
  \bottomrule
  \multicolumn{5}{l}{\textit{Note:} $^{*} p < 0.05, ^{**} p < 0.01, ^{***} p < 0.001$. Standard errors in parentheses.}
\end{tabular}
\end{table}

\begin{table}[htbp]
\CAREtablestyle
  \caption{Blockwise regression results for the Imgflip dataset.}
  \label{tab:imgflip_full_regression}
\begin{tabular}{lcccc}
  \toprule
  \textbf{Variables} & \textbf{Model 1} & \textbf{Model 2} & \textbf{Model 3} & \textbf{Model 4} \\
  \midrule
  (Constant) & $1.164786^{***}$ & $1.178388^{***}$ & $1.045623^{***}$ & $1.040071^{***}$ \\
  \midrule
  \textbf{\textit{Control Variables}} & & & & \\
  \quad \textit{Content Length} & $0.002793^{***}$ & $0.002835^{***}$ & $0.002731^{***}$ & $0.002734^{***}$ \\
  \midrule
  \textbf{\textit{Affective Signals}} & & & & \\
  \quad \textit{Textual Sentiment} &  & $-0.003111$ & $-0.001741$ & $-0.002759$ \\
  \quad \textit{Visual Sentiment} &  & $0.007889^{*}$ & $0.007802^{*}$ & $0.008044^{*}$ \\
  \quad \textit{Happy} &  & $-0.019243^{***}$ & $-0.015080^{***}$ & $-0.014519^{***}$ \\
  \quad \textit{Calm} &  & $-0.027662^{***}$ & $-0.027675^{***}$ & $-0.027596^{***}$ \\
  \quad \textit{Angry} &  & $-0.065530^{***}$ & $-0.059197^{***}$ & $-0.060537^{***}$ \\
  \quad \textit{Sad} &  & $-0.000225$ & $0.005737$ & $0.005017$ \\
  \quad \textit{Fear} &  & $0.001966$ & $0.005863$ & $0.004976$ \\
  \quad \textit{Disgust} &  & $0.028858^{**}$ & $0.031069^{***}$ & $0.029836^{***}$ \\
  \quad \textit{Surprise} &  & $-0.016168^{**}$ & $-0.011618^{*}$ & $-0.011869^{*}$ \\
  \midrule
  \textbf{\textit{Cross-modal Consistency}} & & & & \\
  \quad \textit{Semantic Consistency} &  &  & $0.259351^{***}$ & $0.259850^{***}$ \\
  \quad \textit{Affective Consistency} &  &  &  & $0.006376$ \\
  \midrule
  \textbf{\textit{$R^2$}} & 0.109721 & 0.112672 & 0.117928 & 0.117970 \\
  \textbf{\textit{Adjusted $R^2$}} & 0.109706 & 0.112524 & 0.117766 & 0.117793 \\
  \bottomrule
  \multicolumn{5}{l}{\textit{Note:} $^{*} p < 0.05, ^{**} p < 0.01, ^{***} p < 0.001$. Coefficients are reported.}
\end{tabular}
\end{table}

Tables~\ref{tab:douban_full_regression} and \ref{tab:imgflip_full_regression} report the regression estimates. For Douban, all four models use 13,387 observations. Adjusted $R^2$ increases from 0.051882 with controls to 0.056656 after sentiment and facial-expression indicators are added, to 0.059659 after semantic consistency is added, and to 0.065618 in the full model. For Imgflip, adjusted $R^2$ increases from 0.109706 with controls to 0.112524 with sentiment and facial-expression indicators, to 0.117766 with semantic consistency, and to 0.117793 in the full model. 

Semantic consistency has a positive, statistically significant coefficient in both reported full models (Douban: 0.9354; Imgflip: 0.259850). Affective consistency is positive and statistically significant in Douban ($\beta = 0.2414$, $p < 0.001$), but its Imgflip coefficient is 0.006376 and is not statistically significant. On Imgflip, adding affective consistency changes $R^2$ from 0.117928 to 0.117970, an increase of 0.000042. Textual sentiment is not statistically significant in the full Imgflip model, and several emotion coefficients differ across platforms. These patterns motivate testing affective representations in retrieval, but they do not establish a universal preference for emotional consistency. The regressions also do not evaluate query--response compatibility: their consistency variables describe relationships within a meme.

\section{CARE: Conversational Affective Response Retrieval Framework}

\label{framework}

\begin{figure}[htbp]
  \centering
  \includegraphics[width=\linewidth]{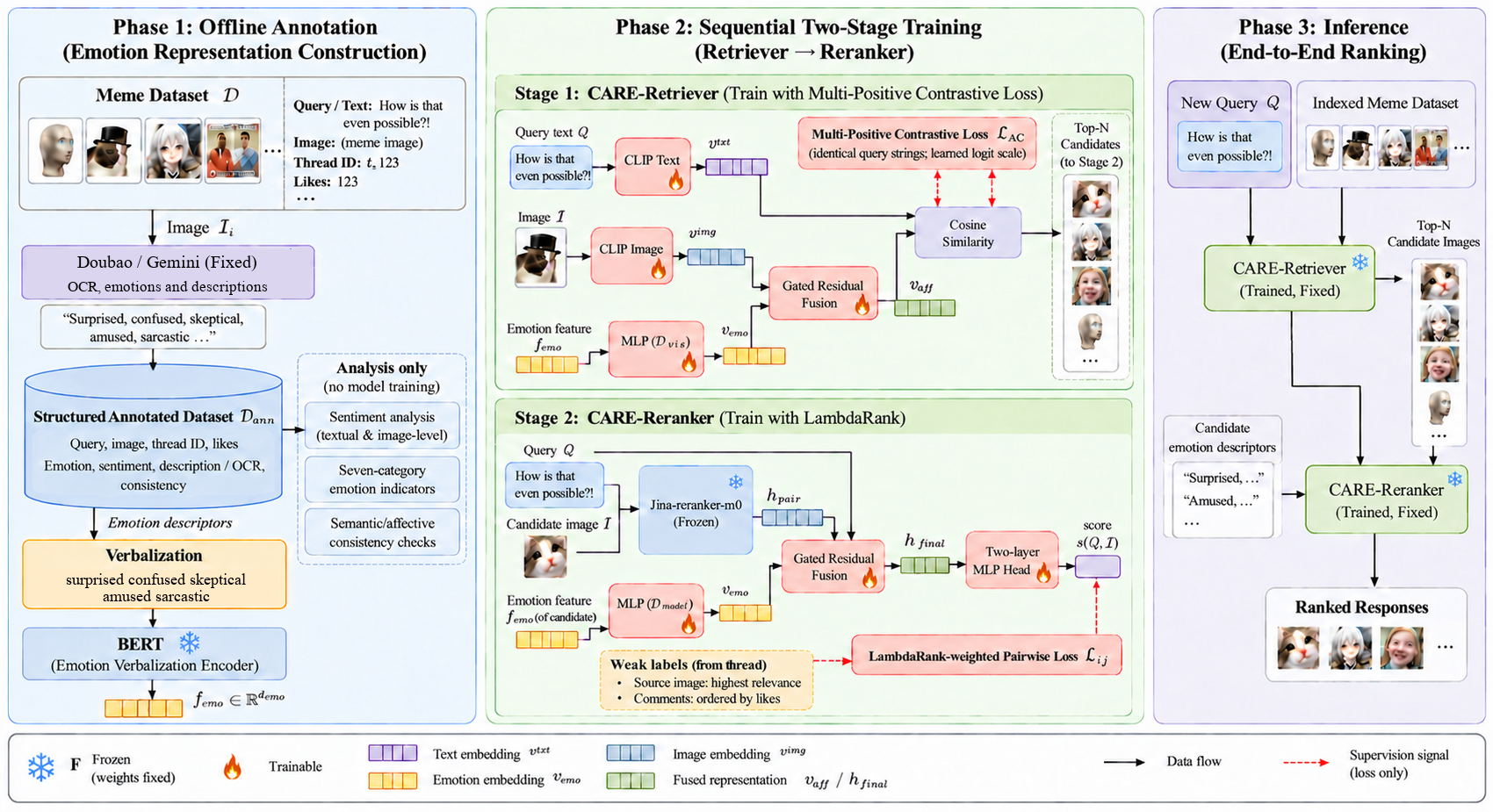}
  \caption{Overview of CARE (Conversational Affective Response Retrieval). Phase 1 assembles raw records and automatic annotations into $D_{\mathrm{ann}}$, then encodes emotion descriptors with frozen BERT. Phase 2 trains the emotion-enabled retriever and then the reranker. Phase 3 ranks candidates with all parameters fixed; the reranker receives the original query and each candidate's image and emotion descriptors. Sentiment and consistency variables support exploratory analysis. Source-image priority applies to eligible source images present in the training candidate set.}
  \label{fig:architecture}
\end{figure}

We propose \textbf{CARE} (Conversational Affective Response Retrieval), a two-stage framework (Figure~\ref{fig:architecture}). CARE-Retriever augments CLIP image embeddings with automatically extracted emotion descriptors. CARE-Reranker combines query--image interactions with emotion features and uses engagement-based ranking supervision.

Figure~\ref{fig:architecture} organizes the workflow into offline annotation and dataset construction (Phase 1), sequential retriever and reranker training (Phase 2), and inference with fixed parameters (Phase 3). In Phase 1, raw records and automatically derived annotations are assembled into a structured annotated collection $D_{\mathrm{ann}}$. Each record retains its query text, image reference, thread membership, source/comment role, and engagement metadata, together with visual descriptions, available OCR text, character emotion descriptors, sentiment labels, and consistency measures. Query--image pairs and emotion descriptors provide model inputs, while thread associations and engagement metadata support weak supervision. The additional sentiment and consistency variables are retained for exploratory analysis. The thread-organized training view $D$ in Algorithm~\ref{alg:care} is derived from $D_{\mathrm{ann}}$; the two symbols denote different views of the same annotated collection.

The architecture uses the richer free-form emotion descriptors as affective inputs, rather than the seven binary facial-expression indicators used in regression. Textual sentiment, image-level polarity, and the semantic and affective consistency scores used in regression are not separate inputs or loss terms in the formulation below. Figure~\ref{fig:architecture} depicts the emotion-enabled branch; the released repository also supports configurations without emotion fusion.

\begin{algorithm}[hbtp]
\caption{CARE Training Pipeline}
\label{alg:care}
\textbf{Input:} Conversational meme dataset $D=\{(q_i, m_{q_i}, M_{q_i}, Y_{q_i}, A_{q_i})\}$, CLIP model, Jina-reranker-m0 model \\
\textbf{Output:} Trained CARE-Retriever and CARE-Reranker
\par\smallskip {\raggedright $Y$: weak ranking labels; $A$: emotion descriptors.\par}
\begin{algorithmic}[1]
\STATE \textbf{Stage 1: CARE-Retriever Training}
\FOR{batch of query--image training pairs $(Q_k,I_k,A_k)$ derived from $D$}
    \STATE Extract visual features $v_{\text{img}}$ from $\{I_k\}$ and text features $v_{\text{txt}}$ from $\{Q_k\}$ using CLIP
    \STATE Encode emotions $\{A_k\}$ via BERT and project to $v_{\text{emo}}$
    \STATE Fuse: $v_{\text{aff}} \leftarrow v_{\text{img}} + \text{Gate}(v_{\text{img}}, v_{\text{emo}}) \odot v_{\text{emo}}$
    \STATE Compute similarity matrix $S_{i,j} = \langle v_{\text{aff}}^{(i)}, v_{\text{txt}}^{(j)} \rangle$
    \STATE Compute Multi-Positive Loss $\mathcal{L}_{\text{AC}}$ using positive pairs
    \STATE Update CLIP and emotion adapter parameters
\ENDFOR
\STATE \textbf{Stage 2: CARE-Reranker Training}
\FOR{batch of query--image training pairs $(Q_k,I_k,A_k)$ derived from $D$}
    \STATE Retrieve top candidates for each query $Q_k$ using the trained retriever
    \STATE Compute cross-encoder features $h_{pair}$ for query-candidate pairs $(Q_k, I)$
    \STATE Encode each candidate's emotion descriptors and project to $v_{\text{emo}} \in \mathbb{R}^{D_{model}}$
    \STATE Fuse emotion: $h_{final} \leftarrow h_{pair} + \text{Gate}(h_{pair}, v_{\text{emo}}) \odot v_{\text{emo}}$
    \STATE Compute relevance scores $s(Q_k, I)$ via scoring head
    \STATE Compute LambdaRank loss $\mathcal{L}_{ij}$ weighted by $|\Delta NDCG_{ij}|$
    \STATE Update scoring-head and emotion-adapter parameters (backbone frozen)
\ENDFOR
\end{algorithmic}
\end{algorithm}

\subsection{CARE-Retriever: Affective Candidate Retrieval}

The CARE-Retriever is implemented as an affect-enhanced CLIP module.

\subsubsection{Emotion Verbalization and Encoding}
Given a meme image $I$ with free-form emotion descriptors $A_I=\{a_1,\dots,a_k\}$, we verbalize $A_I$ into a text descriptor $T_{\text{emo}}$ by concatenating the supplied nonempty emotion descriptors  (e.g., "happy angry"). This concatenation strategy preserves the raw affective signals without additional natural-language templates.

We encode $T_{\text{emo}}$ using frozen \texttt{bert-base-chinese} \citep{devlin2019bert}, taking its pooler output to obtain $f_{\text{emo}} \in \mathbb{R}^{D_{bert}}$, and project it to the CLIP visual space with a two-layer MLP:
\begin{equation}
    v_{\text{emo}} = W_2(\text{ReLU}(W_1 f_{\text{emo}} + b_1)) + b_2
\end{equation}
where $v_{\text{emo}} \in \mathbb{R}^{D_{vis}}$ and $D_{vis}$ denotes the shared CLIP retrieval embedding dimension.

\subsubsection{Gated Affective Fusion}
Let $v_{\text{img}} \in \mathbb{R}^{D_{vis}}$ be the CLIP image embedding. A learnable gate controls fusion of $v_{\text{emo}}$ into $v_{\text{img}}$.

The gate $g \in \mathbb{R}^1$ is computed via a \textbf{Gating Network} as:
\begin{equation}
    g = \sigma(W_g [v_{\text{img}}; v_{\text{emo}}] + b_g)
\end{equation}
where $[\cdot; \cdot]$ denotes concatenation and $\sigma$ is the sigmoid function. The final affect-enhanced visual embedding $v_{\text{aff}}$ is given by:
\begin{equation}
    v_{\text{aff}} = v_{\text{img}} + g \odot v_{\text{emo}}
\end{equation}

\subsubsection{Multi-Positive Contrastive Learning}
Since queries can map to multiple images, we use a multi-positive InfoNCE loss. Given a batch of $N$ image--text pairs $\{(v_{\text{aff}}^{(i)}, v_{\text{txt}}^{(i)})\}_{i=1}^N$, let $S_{i,j} = \langle v_{\text{aff}}^{(i)}, v_{\text{txt}}^{(j)} \rangle$ denote cosine similarity, with $\langle u,v\rangle := u^{\top}v/(\|u\|\|v\|)$. For each anchor image $i$, let $P_i = \{j \mid Q^{(j)} = Q^{(i)}\}$ be the indices of positive texts. The loss is:
\begin{equation}
    \mathcal{L}_{\text{AC}} = \frac{1}{2}\Big(\mathcal{L}_{v \to \text{txt}}^{\mathrm{MP}} + \mathcal{L}_{\text{txt} \to v}^{\mathrm{MP}}\Big)
\end{equation}
where $\mathcal{L}_{v \to \text{txt}}^{\mathrm{MP}}$ is the multi-positive loss for the image-to-text direction:
\begin{equation}
    \mathcal{L}_{v \to \text{txt}}^{\mathrm{MP}} = -\frac{1}{N} \sum_{i=1}^N \frac{1}{|P_i|} \sum_{j \in P_i} \log \frac{\exp(S_{i,j} / \tau)}{\sum_{k=1}^N \exp(S_{i,k} / \tau)}
\end{equation}
Similarly, for the text-to-image direction, let $V_j = \{i \mid Q^{(i)} = Q^{(j)}\}$ be the set of positive images for text $j$. The loss is:
\begin{equation}
    \mathcal{L}_{\text{txt} \to v}^{\mathrm{MP}} = -\frac{1}{N} \sum_{j=1}^N \frac{1}{|V_j|} \sum_{i \in V_j} \log \frac{\exp(S_{i,j} / \tau)}{\sum_{k=1}^N \exp(S_{k,j} / \tau)}
\end{equation}
where $\tau$ is the temperature hyperparameter. In the released implementation, positives are matched by exact equality of query strings, rather than by explicit thread identifiers. Thus, identical query strings from different threads are also treated as positives. The inverse temperature is implemented through CLIP's learned exponential logit scale.

\subsection{CARE-Reranker: Engagement-Supervised Response Ranking}

\label{subsec:rerank}

The CARE-Reranker is a Jina-based cross-encoder that scores query--meme response pairs.

\subsubsection{Query-Image Cross-Encoding}
Given a query $Q$ and candidate image $I$, the JinaVL backbone extracts the final-layer, last-token query--image representation $h_{pair} \in \mathbb{R}^{D_{model}}$:
\begin{equation}
    h_{pair} = \text{JinaVL}(Q, I)
\end{equation}

\subsubsection{Gated Affective Fusion}
Let $v_{\text{emo}} \in \mathbb{R}^{D_{model}}$ be the emotion embedding projected to the Jina feature space (using the same encoding-and-projection construction as in CARE-Retriever, with projection output dimension $D_{model}$). We fuse $v_{\text{emo}}$ into $h_{pair}$ via a learnable gate.

The gate $g \in \mathbb{R}^1$ is computed via a \textbf{Gating Network} as:
\begin{equation}
    g = \sigma(W_r [h_{pair}; v_{\text{emo}}] + b_r)
\end{equation}
where $[\cdot; \cdot]$ denotes concatenation and $\sigma$ is the sigmoid function. The final affect-enhanced representation $h_{final}$ is given by:
\begin{equation}
    h_{final} = h_{pair} + g \odot v_{\text{emo}}
\end{equation}

The final relevance score is computed by a two-layer MLP Scoring Head:
\begin{equation}
    h_{score} = \text{ReLU}(W_{score}^{(1)} h_{final} + b_{score}^{(1)})
\end{equation}
\begin{equation}
    s(Q, I) = W_{score}^{(2)} h_{score} + b_{score}^{(2)}
\end{equation}
where $W_{score}^{(1)} \in \mathbb{R}^{D_{model} \times D_{model}}$, $W_{score}^{(2)} \in \mathbb{R}^{1 \times D_{model}}$ are the weight matrices, and $b_{score}^{(1)}$, $b_{score}^{(2)}$ are bias terms.

\subsubsection{LambdaRank Optimization}
\label{subsec:lambdarank}
The source meme image is assigned the highest relevance label, taking precedence over the reciprocal-rank labeling rule. For the remaining candidates in each training candidate list, candidates with matching query text and positive like counts are ordered by decreasing numbers of likes. A candidate at positive rank $r$ receives a relevance label of $y=1/r$, while all other candidates receive a label of zero. Ties in like counts retain the original candidate order. The NDCG gain associated with a label $y$ is defined as $2^y-1$.

To optimize the ranking order, we use a LambdaRank-weighted pairwise logistic objective \citep{NIPS2006_af44c4c5}. For a pair of documents $d_i$ and $d_j$ where $d_i$ is more relevant than $d_j$ (i.e., $y_i > y_j$), the gradients are scaled by the change in NDCG ($\Delta NDCG$) that would result from swapping their positions.

The pairwise loss is weighted by $\Delta NDCG$:
\begin{equation}
    \mathcal{L}_{ij} = - |\Delta NDCG_{ij}| \log \frac{\exp({\scriptstyle \gamma} s_i)}{\exp({\scriptstyle \gamma} s_i) + \exp({\scriptstyle \gamma} s_j)}
\end{equation}
where $s_i$ and $s_j$ are the predicted scores, and $\gamma$ is a scaling parameter. The weighting emphasizes swaps with larger effects on the chosen ranking metric. It relies on the weak relevance labels defined above.

\section{Experimental Evaluation and Analysis}\label{experiments}

\subsection{Evaluation Protocol}

The Douban evaluation uses 885 query records randomly sampled from the test split, while the Imgflip evaluation uses 1,000 query records. The full candidate pools contain 4,108 test images for Douban and 15,742 for Imgflip. During reranking, the reranker reorders the top 100 candidates returned by the first-stage retriever. Results are averaged over the sampled query records.

Each query corresponds to the textual context of a main post and may be associated with multiple thread-associated meme records, including the source meme and comment memes in the same thread. Therefore, we adopt a \textbf{multi-positive evaluation protocol}. Unlike one-to-one image--text retrieval, conversational meme retrieval is open-ended: a retrieval is considered successful if at least one meme from the ground-truth response set $G_q$ is retrieved in the top-$K$ results.

We report Hit@K ($H@K$), the proportion of queries with at least one labeled positive in the top-$K$. This differs from multi-positive recall, which measures the fraction of all relevant items recovered:

\begin{align}
    H@K &= \frac{1}{|Q|} \sum_{q \in Q} \mathbb{I}(G_q \cap R_{q,K} \neq \varnothing) \\
    P@K &= \frac{1}{|Q|} \sum_{q \in Q} \frac{|G_q \cap R_{q,K}|}{K}
\end{align}
where $Q$ is the test query set and $G_q=\{m_q\}\cup M_q$ is the thread-associated positive set for query $q$; $R_{q,K}$ denotes the top-$K$ retrieved candidates, and $\mathbb{I}(\cdot)$ is the indicator function.

Additionally, to evaluate whether the model can recover the original visual context of the conversation, we define \textbf{Main-Post Recall@K} ($MR@K$):
\begin{equation}
    MR@K = \frac{1}{|Q|} \sum_{q \in Q} \mathbb{I}(I_q \in R_{q,K})
\end{equation}
where $I_q=m_q$ denotes the source meme image attached to the main post.

\subsection{Dataset Splitting}
To evaluate generalization across conversational contexts, we split the benchmark by \textbf{topic} rather than by individual comment. Douban contains \textbf{951 training topics} and \textbf{885 test topics}, totaling 1,836 topics; the test proportion is 48.20\%. Imgflip contains \textbf{2,997 training topics} and \textbf{750 test topics}, totaling 3,747 topics; the test proportion is 20.02\%.

\subsection{Implementation Details}

CARE uses \textbf{Chinese-CLIP} \citep{yang2022chinese} for candidate retrieval on Douban and \textbf{CLIP ViT-B/32} \citep{Radford2021LearningTV} on Imgflip, with \texttt{jina-reranker-m0} for fine-grained response ranking.

\begin{itemize}

    \item \textbf{CARE-Retriever Training:} The retriever is fine-tuned for 5 epochs with a batch size of 128. We use Adam with a shared learning rate of $5\times 10^{-6}$ for the pretrained retrieval backbone and the affective fusion parameters. The emotion extraction module and BERT parameters are frozen; the emotion projection and scalar gate are trainable.

    \item \textbf{CARE-Reranker Training:} The reranker freezes the pretrained cross-encoder backbone and updates only the scoring head and emotion adapter. It is trained with Adam using a learning rate of $1\times 10^{-6}$ and the LambdaRank objective \citep{NIPS2006_af44c4c5} for up to 10 epochs. Numerical relevance labels are constructed as described in Section~\ref{subsec:lambdarank}. The source meme is assigned the highest relevance label to preserve its role as the original visual context. The head-only ablation trains the Jina scoring head without emotion features.

\end{itemize}

\subsection{Baselines}
We compare our method against the following retrieval baselines:
\begin{itemize}
    \item \textbf{VSE++ \citep{faghri2018vse++}:} A classic approach learning a common embedding space with hard negative mining.
    \item \textbf{CLIP \citep{Radford2021LearningTV}:} Evaluated in both zero-shot and fine-tuned settings, using Chinese-CLIP on Douban and CLIP ViT-B/32 on Imgflip.
    \item \textbf{BLIP \citep{li2022blip}:} A unified vision-language pre-training framework with strong retrieval capabilities.
    \item \textbf{IEF \citep{10.1117/12.3065281}:} An emotion-infused multimodal recommendation framework.
\end{itemize}
The baselines use image and text inputs, while IEF and CARE additionally use affective information. The complete CARE system also includes a trained reranker. Consequently, the main comparison evaluates complete systems rather than isolating emotion-feature fusion under matched architecture and training.

\subsection{Main Results}
Table \ref{tab:main_results} presents the retrieval performance on the Douban (Chinese) and Imgflip (English) test sets. Here, \textbf{CARE} denotes our complete two-stage framework. All values are reported as percentages (\%), and the best results are highlighted in \textbf{bold}.

\begin{table}[!htbp]
\CAREtablestyle
  \centering
  \caption{Main results on the Douban and Imgflip test sets.}
  \label{tab:main_results}
\begin{tabular}{llcccccccc}
    \toprule
    \textbf{Dataset} & \textbf{Model} & \textbf{H@1} & \textbf{H@3} & \textbf{H@5} & \textbf{H@10} & \textbf{H@50} & \textbf{P@3} & \textbf{P@5} & \textbf{P@10} \\
    \midrule
    \multirow{5}{*}{Douban} & VSE++ & 2.3 & 3.2 & 4.1 & 7.5 & 13.45 & 2.6 & 2.0 & 1.5 \\
     & CLIP & 10.40 & 13.8 & 17.74 & 23.16 & \textbf{32.4} & 7.9 & 7.21 & 6.05 \\
     & BLIP & 10.06 & 13.1 & 15.93 & 21.02 & 30.5 & 7.0 & 6.49 & 5.2 \\
     & IEF & 10.6 & 12.9 & 15.14 & 20.0 & 29.94 & 6.8 & 6.0 & 4.8 \\
     & \textbf{CARE} & \textbf{13.63} & \textbf{18.72} & \textbf{22.25} & \textbf{24.03} & 29.70 & \textbf{10.13} & \textbf{8.60} & \textbf{6.72} \\
    \midrule
    \multirow{5}{*}{Imgflip} & VSE++ & 7.07 & 11.33 & 15.60 & 21.33 & 32.00 & 6.18 & 5.01 & 4.29 \\
     & CLIP & 21.20 & 26.5 & 35.87 & 41.07 & 52.93 & 13.0 & 11.12 & 8.16 \\
     & BLIP & 22.27 & 28.9 & 35.33 & 41.20 & 52.0 & 12.5 & 11.04 & 9.5 \\
     & IEF & 16.27 & 22.4 & 28.53 & 33.47 & 48.80 & 10.2 & 9.12 & 7.96 \\
     & \textbf{CARE} & \textbf{25.60} & \textbf{38.27} & \textbf{41.33} & \textbf{49.33} & \textbf{61.33} & \textbf{18.00} & \textbf{13.44} & \textbf{10.77} \\
    \bottomrule
  \end{tabular}
\end{table}

CARE achieves Hit@10 of 24.03\% on Douban and 49.33\% on Imgflip. Relative to the fine-tuned CLIP values in Table~\ref{tab:main_results}, these are absolute increases of 0.87 and 8.26 percentage points, corresponding to relative improvements of 3.8\% and 20.1\%. On Imgflip, CARE also exceeds the strongest non-CARE comparator, BLIP, by 8.13 points at Hit@10. Across the two datasets, CARE attains the best result in fifteen of the sixteen reported metric columns. CARE leads on the reported early-rank metrics, while fine-tuned CLIP retains higher Hit@50 on Douban.

The gains are most pronounced at the top of the ranking, which is the operating range that matters for a response-suggestion system. On Douban, Hit@1 rises from 10.40\% to 13.63\% (a 31.1\% relative improvement) and Precision@3 from 7.9\% to 10.13\% (+28.2\%). On Imgflip, Hit@3 increases from 26.5\% to 38.27\% (+44.4\%) and Precision@10 from 8.16\% to 10.77\% (+32.0\%). Because a deployed meme assistant typically presents only a handful of candidates, lifting the first few ranks has greater practical value than marginal changes at deep cutoffs, and the Precision@K results confirm that the promoted items are themselves labeled positives, evidence that reranking surfaces new positives in the top ranks.

The comparison with IEF evaluates two affect-aware systems. At Hit@10, IEF underperforms fine-tuned CLIP on both collections (for example, Imgflip: 33.47\% versus 41.07\%), whereas CARE reaches 49.33\%. CARE retains the CLIP image embedding as the residual base for gated emotion fusion and also includes supervised reranking. This complete-system comparison does not isolate the contribution of the fusion mechanism or establish why IEF performs worse.

The two collections show different ranking profiles. On Imgflip, CARE improves over fine-tuned CLIP at every cutoff, including Hit@50 (61.33\% versus 52.93\%, a 15.9\% relative gain). On Douban, the largest gains appear at Hit@1--Hit@5, whereas fine-tuned CLIP retains higher Hit@50 (32.4\% versus 29.70\%). The reranker reorders a fixed top-100 candidate list using a LambdaRank objective weighted by changes in NDCG. The ablations below compare the stages: CARE-Retriever alone improves over fine-tuned CLIP in all eight reported columns (Table~\ref{tab:ablation_retrieval}), while adding reranking changes the balance across cutoffs.

\subsection{Ablation Studies}

The ablation studies examine CARE-Retriever and CARE-Reranker. CARE-Retriever denotes the retrieval stage alone; adding CARE-Reranker produces the complete CARE system.

\subsubsection{Effect of CARE-Retriever}
To examine CARE-Retriever, we compare the retrieval stage alone against the platform-specific CLIP backbone in zero-shot and fine-tuned settings. The results are summarized in Table \ref{tab:ablation_retrieval}, where values are percentages (\%) and best results are in \textbf{bold}.
\begin{table}[htbp]
\CAREtablestyle
  \caption{Ablation study of the effect of CARE-Retriever.}  \label{tab:ablation_retrieval}
\begin{tabular}{lcccccccc}
    \toprule
    \multirow{2}{*}{\textbf{Model}} & \multicolumn{4}{c}{\textbf{Douban}} & \multicolumn{4}{c}{\textbf{Imgflip}} \\
    \cmidrule(lr){2-5} \cmidrule(lr){6-9}
    & \textbf{H@1} & \textbf{H@3} & \textbf{H@5} & \textbf{H@10} & \textbf{H@1} & \textbf{H@3} & \textbf{H@5} & \textbf{H@10} \\
    \midrule
    CLIP (zero-shot) & 1.13 & 3.16 & 5.31 & 5.99 & 18.80 & 27.47 & 36.13 & \textbf{42.67} \\
    CLIP (fine-tuned) & 10.40 & 13.8 & 17.74 & 23.16 & 21.20 & 26.5 & 35.87 & 41.07 \\
    CARE-Retriever & \textbf{11.53} & \textbf{15.03} & \textbf{17.97} & \textbf{26.89} & \textbf{25.33} & \textbf{30.93} & \textbf{37.47} & 42.40 \\
    \bottomrule
  \end{tabular}
\end{table}

Table~\ref{tab:ablation_retrieval} shows that CARE-Retriever outperforms fine-tuned CLIP at all evaluated cutoffs on both datasets and achieves the best results in seven of the eight reported metric columns. On Douban, Hit@10 increases from 23.16\% to 26.89\%, an absolute gain of 3.73 percentage points (a 16.1\% relative improvement). On Imgflip, CARE-Retriever achieves Hit@1 of 25.33\% and Hit@3 of 30.93\%, exceeding fine-tuned CLIP by 4.13 and 4.43 percentage points, respectively.

CARE-Retriever combines gated residual fusion with multi-positive contrastive learning. The fusion adds a weighted emotion vector to the CLIP image embedding. The loss treats thread-associated memes as positives, accommodating several labeled candidates for a query. These design choices are evaluated together here; the results do not isolate the contribution of each. On Imgflip, zero-shot CLIP reaches 42.67\% Hit@10, while fine-tuned CLIP reaches 41.07\% and improves Hit@1 from 18.80\% to 21.20\%. CARE-Retriever attains Hit@1 of 25.33\% and Hit@10 of 42.40\%. Thus, its Hit@10 remains close to the zero-shot reference while its early-rank hit rates are higher.

\subsubsection{Effect of CARE-Reranker}
To examine the effect of CARE-Reranker, we compare retrieval alone with two reranking configurations: a trained Jina scoring head without emotion features and CARE-Reranker. All configurations use candidates generated by CARE-Retriever. Table~\ref{tab:ablation_rerank} includes the head-only control, while Table~\ref{tab:rerank_mainpost_results} reports Main-Post Recall for the same three configurations. In both tables, rows prefixed with ``+'' add the indicated reranker to CARE-Retriever.
\begin{table}[htbp]
\CAREtablestyle
  \centering
  \caption{Ablation study of the effect of CARE-Reranker.}
  \label{tab:ablation_rerank}
\begin{tabular}{lcccccccc}
    \toprule
    \multirow{2}{*}{\textbf{Configuration}} & \multicolumn{4}{c}{\textbf{Douban}} & \multicolumn{4}{c}{\textbf{Imgflip}} \\
    \cmidrule(lr){2-5} \cmidrule(lr){6-9}
    & \textbf{H@1} & \textbf{H@3} & \textbf{H@5} & \textbf{H@10} & \textbf{H@1} & \textbf{H@3} & \textbf{H@5} & \textbf{H@10} \\
    \midrule
    CARE-Retriever & 11.53 & 15.03 & 17.97 & \textbf{26.89} & 25.33 & 30.93 & 37.47 & 42.40 \\
    + Jina head (no emotion) & 13.12 & 18.30 & 22.18 & 23.81 & 24.60 & 37.30 & 39.20 & 46.40 \\
    \textbf{+ CARE-Reranker} & \textbf{13.63} & \textbf{18.72} & \textbf{22.25} & 24.03 & \textbf{25.60} & \textbf{38.27} & \textbf{41.33} & \textbf{49.33} \\
    \bottomrule
  \end{tabular}
\end{table}

Table~\ref{tab:ablation_rerank} compares supervised reranking with and without the emotion adapter on the same CARE-Retriever candidates. On Imgflip, CARE-Reranker raises Hit@10 from 42.40\% to 49.33\% (+6.93 points) and Hit@3 from 30.93\% to 38.27\%. Relative to the trained head-only control, the emotion-adapter configuration adds 2.93 points at Hit@10 and 2.13 points at Hit@5. On Douban, its margins over the head-only control are +0.51 points at Hit@1, +0.42 at Hit@3, +0.07 at Hit@5, and +0.22 at Hit@10. These results are consistent with a benefit from the emotion-enhanced reranking configuration. The comparisons do not isolate gating from other adapter changes, and uncertainty estimates across queries or repeated runs are not reported.

Relative to retrieval alone, reranking on Douban lowers Hit@10 from 26.89\% to 24.03\% while raising Hit@1 from 11.53\% to 13.63\%, Hit@3 from 15.03\% to 18.72\%, and Hit@5 from 17.97\% to 22.25\%. This improves early-rank hit rates at the expense of Hit@10. The complete system also has the highest reported Precision@K values in Table~\ref{tab:main_results}. On Imgflip, adding CARE-Reranker improves every reported cutoff. These observations describe different trade-offs between the collections; they do not establish which dataset properties cause those differences.

\paragraph{Main-Post Recall under Two-Stage Reranking}
Table~\ref{tab:rerank_mainpost_results} verifies that reranking improves the rank position of the main-post meme image. This improvement is expected because, as noted in the training details, the main post image is assigned the highest relevance label during reranker training. Consequently, the model is directly supervised to recover the ground-truth post, leading to numerical gains in Main-Post Recall compared to CARE-Retriever alone (e.g., more than doubling MR@10 on Imgflip).
\begin{table}[htbp]
\CAREtablestyle
  \centering
  \caption{Effect of CARE-Reranker on end-to-end Main-Post Recall@K.}
  \label{tab:rerank_mainpost_results}
\begin{tabular}{lcccccccc}
    \toprule
    \multirow{2}{*}{\textbf{Setting}} & \multicolumn{4}{c}{\textbf{Douban (Chinese)}} & \multicolumn{4}{c}{\textbf{Imgflip (English)}} \\
    \cmidrule(lr){2-5} \cmidrule(lr){6-9}
    & \textbf{MR@1} & \textbf{MR@3} & \textbf{MR@5} & \textbf{MR@10} & \textbf{MR@1} & \textbf{MR@3} & \textbf{MR@5} & \textbf{MR@10} \\
    \midrule
    CARE-Retriever & 1.1 & 1.9 & 2.5 & 4.3 & 0.7 & 2.0 & 2.8 & 3.6 \\
    + Jina head (no emotion) & 2.5 & 2.94 & 4.2 & 7.7 & 2.1 & 2.93 & 4.3 & 5.33 \\
    \textbf{+ CARE-Reranker} & \textbf{2.7} & \textbf{3.5} & \textbf{5.6} & \textbf{8.2} & \textbf{2.53} & \textbf{3.3} & \textbf{5.73} & \textbf{7.73} \\
    \bottomrule
  \end{tabular}
\end{table}

Concretely, MR@1 rises from 1.1\% to 2.7\% on Douban and MR@10 from 3.6\% to 7.73\% on Imgflip, and the emotion-equipped configuration outperforms the head-only control in all eight reported columns. Under the topic-held-out protocol, the reranker scores query--image pairs together with emotion descriptors. The results show improved source-image recovery under LambdaRank supervision combining likes with source-image priority. They do not establish conversational reply quality or rule out reliance on source-image cues.

\subsection{Qualitative Analysis of Affective Retrieval}
\begin{figure}[htbp]
  \centering
  \begin{subfigure}{\linewidth}
    \centering
    \includegraphics[width=0.8\linewidth]{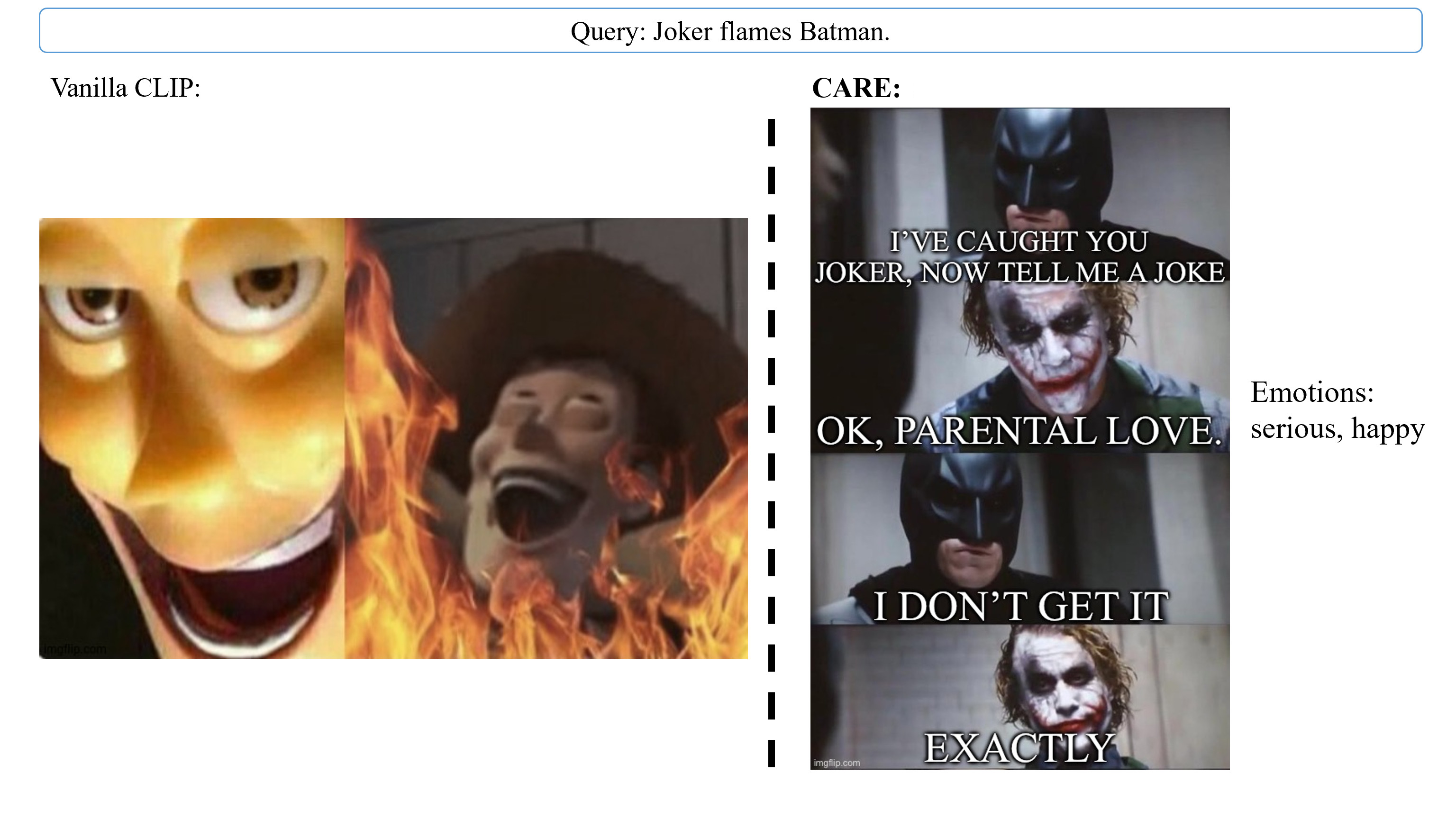}
    \caption{Comparison of retrieval results for the query ``Joker flames Batman'' on the Imgflip dataset.}
    \label{fig:case_study_a}
  \end{subfigure}
  \begin{subfigure}{\linewidth}
    \centering
    \includegraphics[width=0.8\linewidth]{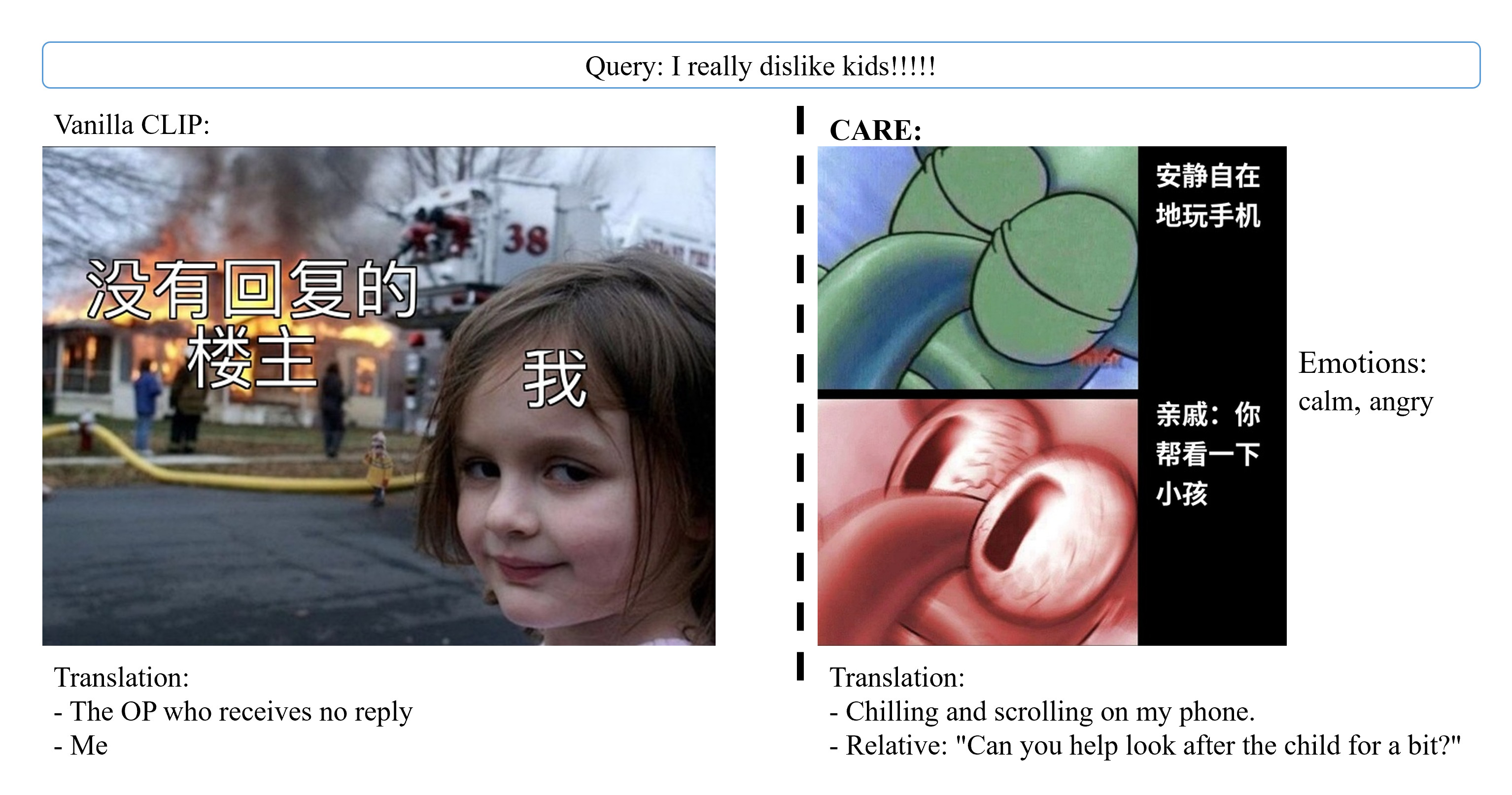}
    \caption{Comparison of retrieval results for the query ``I really dislike kids!!!!!'' on the Douban dataset.}
    \label{fig:case_study_b}
  \end{subfigure}
  \caption{Selected retrieval examples from Imgflip and Douban, comparing CLIP with the complete CARE system.}
  \label{fig:case_studies}
\end{figure}

Figure~\ref{fig:case_studies} presents two selected retrieval examples. For ``Joker flames Batman,'' CLIP retrieves flame imagery, whereas CARE returns a dialogue meme conveying mockery. For ``I really dislike kids!!!!!,'' the CARE result depicts a transition from calmness to anger that appears compatible with the query's expressed frustration. These examples illustrate possible differences between the complete systems. They do not independently identify the effect of emotion features or provide systematic judgments of conversational appropriateness.

\subsection{Cross-Platform Synthesis: From Engagement Signals to Retrieval Behavior}\label{subsec:synthesis}
The exploratory regression and the retrieval experiments target different outcomes: variation in log-likes and thread-associated ranking, respectively. Engagement is the regression outcome and also contributes to reranker supervision. Their findings can inform the retrieval design, but the regression is not an independent validation of the retrieval mechanism.

First, semantic consistency is positively associated with engagement in both full regression models (Douban: 0.9354; Imgflip: 0.2599, both $p<0.001$), and adding the cross-modal blocks raises adjusted $R^2$ from 0.0519 to 0.0656 on Douban and from 0.1097 to 0.1178 on Imgflip. In the retrieval experiments, the strongest comparators are semantic vision--language models. CARE uses the CLIP embedding as a residual input and adds a weighted emotion representation. These findings motivate retaining semantic information alongside affective features; they do not independently validate the gate or guarantee preservation of semantic neighborhoods.

Second, the reported affective associations differ between collections. Affective consistency is significant on Douban ($\beta=0.2414$, $p<0.001$) but not on Imgflip ($\beta=0.0064$, n.s.); its base rates are 0.50 and 0.82, respectively. Individual emotion coefficients also differ, with Disgust positive on Douban (0.3128) and Angry negative on Imgflip ($-0.0605$). The reranker's numerical margins over the head-only control likewise differ between collections. These are different affective channels: facial-expression indicators and within-meme consistency in regression, and richer emotion descriptors in retrieval. Their results should not be mapped one-to-one or interpreted as proof that the gate adapts optimally to platform differences.

Dataset structure also matters when interpreting absolute scores. The Imgflip candidate pool is 3.83 times larger (15,742 versus 4,108 images), and its threads contain 22.72 memes on average versus 7.29 on Douban. Under the evaluation protocol, hit rates depend on both candidate-pool composition and the size of each positive set $G_q$. Language, platform conventions, collection period, and thread structure vary together here. The two collections therefore provide evaluations in different settings, without isolating the causes of their performance differences or demonstrating cross-platform transfer.

\section{Discussion and Limitations}\label{discussion}
\subsection{Principal Findings}
The study yields four main observations. First, under the topic-held-out protocol, the complete CARE system leads in fifteen of sixteen reported metric columns, with a Douban Hit@50 trade-off. Second, CARE-Retriever exceeds fine-tuned CLIP at all evaluated cutoffs, and CARE-Reranker exceeds the trained head-only control in both collections. These comparisons support the evaluated configurations without isolating every component. Third, LambdaRank supervision combining likes and source-image priority improves Main-Post Recall; on Imgflip, MR@10 more than doubles relative to retrieval alone. Fourth, the exploratory affective associations and the reranking margins differ between collections. Their agreement motivates further study of affect-aware retrieval rather than establishing a universal emotion rule.

\subsection{Implications for Affect-Aware Retrieval Design}
The results suggest design considerations for further evaluation. A semantic retrieval backbone can be combined with a weighted affective representation, as in CARE's residual fusion. Multi-positive learning accommodates the thread-based label structure, although thread membership is incomplete relevance supervision. The two-stage design restricts cross-encoding to the top 100 candidates; the Douban results illustrate that gains at early ranks can accompany losses at deeper cutoffs. Likes provide scalable ranking labels, but exposure and audience effects limit their interpretation as measures of response quality.

\subsection{Practical Implications and Research Resources}
CARE offers a potential design for response-suggestion applications. The retrieval backbone is fine-tuned together with the emotion projection and gate, while the emotion extraction module and BERT remain frozen. During reranker training, the pretrained cross-encoder backbone is frozen, and only its scoring head and emotion adapter are updated. Restricting cross-encoding to the top-100 candidates limits the number of reranker evaluations per query; latency, memory, and throughput are not measured here. Applying this pipeline to stickers, emoji, or short-video replies would require further validation. The benchmark provides 98,506 thread-organized records across 5,583 topics and two languages, with automatic affective annotations, engagement labels, and a topic-held-out evaluation protocol.

\subsection{Semantic Relevance and Affective Compatibility}
Emotion similarity is one component of relevance and is not sufficient on its own. Memes expressing the same emotion can concern unrelated events, and an appropriate reply may convey a different emotion from the query, as in reassurance or irony. CARE combines the semantic embedding with an emotion representation and uses query--image interactions in reranking. Further work could examine calibrated context and affect scores, query-conditioned fusion, and hard negatives that share emotion but differ in context. These extensions could help distinguish semantic relevance from affective compatibility.

\subsection{Limitations}
\paragraph{Weak relevance and engagement labels}
Thread membership is incomplete supervision, and likes are influenced by exposure and audience composition. The source image is assigned the highest rank by design, so Main-Post Recall gains measure recovery of the conversational anchor rather than superior comment-response selection. The topic-held-out protocol separates training and test discussions, but does not establish isolation of near-duplicate images or shared templates across topics. Human assessment of semantic relevance, affective compatibility, and conversational usefulness, together with response-only evaluation, remains future work.

\paragraph{Automatic annotation}
Retrieval emotion descriptors may draw on captions and contextual cues as well as facial expressions. Regression instead uses seven facial-expression indicators, together with sentiment and consistency variables. Both kinds of automatic annotation can contain errors. Naming the models and releasing the zero-shot prompts in~\ref{app:llm_prompts} makes the annotation procedure more transparent and auditable, but does not guarantee deterministic outputs or annotation accuracy. An independent human annotation audit remains future work.

\paragraph{Platform and statistical scope}
Douban and Imgflip differ in language, platform conventions, and collection period, so the separate evaluations do not demonstrate cross-lingual transfer or isolate cultural effects. The exploratory blockwise regressions do not account for topic-level dependence or identify causal effects. The results describe associations and ranking performance within the two reported settings; their generalization to other platforms and periods remains to be evaluated.

\subsection{Future Research Directions}
Four directions follow directly from this study. The first is human-centered validation: collecting graded judgments of semantic relevance, affective compatibility, and conversational usefulness, and adding response-only evaluation that excludes the source meme. The second is architectural refinement along the path identified above: query-conditioned gates, disentangled context/affect scores, and emotion-sharing hard negatives. The third is cross-platform and cross-lingual transfer, including shared multilingual encoders and explicit modeling of cultural affect differences. The fourth broadens the learning signal: temporal drift in meme usage, larger platform networks, multimodal large language models as rerankers, and quasi-experimental designs that move the engagement analysis from association toward causation.

\section{Conclusion}\label{conclusion}
We presented CARE, a two-stage conversational meme retrieval framework combining gated emotion descriptors, multi-positive contrastive learning, and engagement-supervised reranking. Under the reported topic-held-out protocol, CARE leads in fifteen of sixteen metric columns and achieves Hit@10 of 24.03\% on Douban and 49.33\% on Imgflip. Its largest gains occur at early ranks, while fine-tuned CLIP retains higher Hit@50 on Douban. The trained head-only comparison is consistent with an additional benefit from the emotion-adapter configuration. The current experiments do not isolate every architectural contribution or establish perceived conversational usefulness.

The thread-organized benchmark and evaluated pipeline provide a basis for further research on affect-aware response retrieval. Human judgments of relevance and usefulness, matched complete-system comparisons, query-conditioned fusion, and cross-lingual transfer are directions for future validation.

\section*{CRediT authorship contribution statement}
\textbf{Jiajun Li}: Data curation, Methodology, Investigation, Formal analysis, Visualization, Writing -- Original Draft, Writing -- Review \& Editing.

\textbf{Leihan Zhang}: Conceptualization, Methodology, Formal analysis, Writing -- Review \& Editing, Project administration, Funding acquisition.

\textbf{Teng Zhang}: Data curation, Investigation, Formal analysis.

\textbf{Le Zhang}: Methodology, Supervision, Writing -- Review \& Editing, Funding acquisition.

\textbf{Qiang Yan}: Project administration, Writing -- Review \& Editing.

\section*{Funding}
This research was supported by the National Natural Science Foundation of China (NSFC) under Grant Nos. 62102044 and 72501034.

\section*{Data availability}
The source code is available at \url{https://github.com/TiffanyBlews/CARE}. The dataset is available at \url{https://huggingface.co/datasets/TiffanyBlews/CARE_Meme_Dataset/tree/main}, subject to the access conditions specified on Hugging Face.

\section*{Declaration of competing interest}
The authors declare that they have no known competing financial interests or personal relationships that could have appeared to influence the work reported in this paper.

\appendix

\section{LLM Prompts for Affective Meme Extraction}\label{app:llm_prompts}
Doubao Vision Pro is used for OCR, emotion annotation, and visual descriptions on Douban, while Gemini Pro is used for these tasks on Imgflip.
\begin{lstlisting}
Please use English to analyze the provided meme image and output the results in JSON format containing the following seven parts:

1. Visual Description (description): Provide a detailed textual description of the visual content of the meme, without providing interpretations.

2. Character Emotions (emotions, emotion_labels): In the same annotation pass, provide two distinct character-level fields: a free-form emotion descriptor for retrieval and a facial-expression label for regression:

- emotions: A list of concise English emotion descriptors inferred from the full meme context for retrieval. These descriptors are not restricted to a predefined vocabulary and may capture nuanced reactions such as amused, skeptical, sarcastic, disappointed, or worried.

- emotion_labels: A list of facial-expression labels for regression, selected only from these seven lowercase English categories: happy, calm, angry, sad, fear, disgust, surprise. Select the single best-fitting category based on each character's facial expression; this label is not a categorical equivalent of the richer contextual emotion descriptor. The label calm denotes a neutral or relaxed facial expression and must not be used as a fallback for uncertainty.

Keep the two lists aligned in reading order: entries at the same index must refer to the same character appearance. Retain repeated descriptors and labels for different character appearances. If no character emotion can be identified, return an empty list for both fields.

3. Image Sentiment Polarity (image_sentiment): Indicates the overall emotional inclination expressed by the meme image. Values are:
   - 1: Positive sentiment
   - 0: Neutral sentiment
   - -1: Negative sentiment

4. Theme (theme): Explain the theme or core idea that this meme image intends to convey.

5. Text Sentiment Polarity (text_sentiment): Indicates the emotional inclination conveyed by the meme text, with the same values as above:
   - 1: Positive sentiment
   - 0: Neutral sentiment
   - -1: Negative sentiment

6. Usage Scenarios (usage_scenarios): Describe the scenarios in which this meme image is typically used.

7. OCR Text (ocr_text): Transcribe the readable text in the image in its original language, preserving line breaks. Do not translate or invent text. Return an empty string if no text is readable.

Please ensure the output strictly follows the following JSON structure:
{
  "description": "...",
  "emotions": ["amused", "worried"],
  "emotion_labels": ["happy", "fear"],
  "image_sentiment": 1,
  "theme": "...",
  "text_sentiment": 0,
  "usage_scenarios": "...",
  "ocr_text": "..."
}
Output only the pure JSON result without any additional explanation.
\end{lstlisting}
Note that while the template includes fields for \textit{Theme} and \textit{Usage Scenarios}, these attributes were not utilized in the experiments of this study and are included in the dataset for future research.

\section*{Declaration of generative AI and AI-assisted technologies in the manuscript preparation process}
During the preparation of this work, the authors used ChatGPT, Gemini, and OpenAI Codex to assist with language editing, manuscript restructuring, and consistency checking. The authors are responsible for reviewing and verifying all assisted content before submission and for the accuracy and integrity of the final manuscript. The use of Doubao Vision Pro and Gemini Pro for research annotation is described separately in the Methods.

\bibliographystyle{elsarticle-harv}

\bibliography{references}

\end{document}